\documentclass[preprint,12pt]{elsarticle}

\usepackage{graphicx}
\usepackage{epstopdf}
\usepackage{amsmath}
\usepackage{amssymb,amsfonts}
\usepackage{gensymb}
\usepackage{enumerate}
\usepackage{color}
\usepackage{float}
\usepackage{multicol}
\usepackage{longtable}
\usepackage{caption} 
\usepackage[export]{adjustbox}
\usepackage{url}
\usepackage{tcolorbox}
\usepackage[labelfont=bf,textfont={sl,bf},lofdepth,lotdepth]{subfig}
\DeclareCaptionListOfFormat{colon}{#2:}
\usepackage{fixltx2e}
\usepackage{subfig}
\usepackage{changepage}
\usepackage{tabulary}
\newcolumntype{K}[1]{>{\centering\arraybackslash}p{#1}}

\usepackage{lineno}

\usepackage[top=1in, left=1.25in, right=1in, bottom=1in]{geometry}
\newcommand{\Rc}{\mathcal{R}^i}
\newcommand{\T}{\mathcal{T}}
\newcommand{\J}{\mathcal{J}}
\newcommand{\Pm}{\mathcal{P}}
\newcommand{\I}{\mathcal{I}}
\newcommand{\Oc}{\mathcal{O}}

\journal{arXiv.org}

\begin{document}

\begin{frontmatter}


\title{On the Interaction between Personal Comfort Systems and Centralized HVAC Systems in Office Buildings}
%An energy efficient personal thermal comfort system for office buildings
%A novel control approach for energy efficient HVAC in office buildings using personal comfort systems


\author{Rachel Kalaimani, Milan Jain, Srinivasan Keshav, Catherine Rosenberg}
\fntext[]{R. Kalaimani is with the Department of Electrical Engineering, IIT Madras, India, email:rachel@ee.iitm.ac.in. M. Jain is with IIIT Delhi, India, email:milanj@iiitd.ac.in. C. Rosenberg is with the Department of Electrical and Computer Engineering
 at University of Waterloo, Canada, email: cath@uwaterloo.ca. S. Keshav is with the Cheriton School of Computer Science at University of Waterloo, Canada, email: keshav@uwaterloo.ca.}
 \fntext[]{This research was supported in part by the Natural Sciences and Engineering Research Council of Canada (NSERC) and was performed when the first two authors were at the University of Waterloo.}

\begin{abstract}
Most modern HVAC systems suffer from two intrinsic problems. 
First, inability to  meet diverse comfort requirements of the occupants. Second, heat or cool an entire zone even when the zone is only partially occupied. Both issues can be mitigated by using personal comfort systems (PCS) which bridge the comfort gap between what is provided by a central 
HVAC system and the personal preferences of the occupants.
In recent work, we have proposed and deployed such a system, called SPOT.

We address the question, ``How should an existing HVAC system modify its operation to benefit  the availability of PCS like SPOT?"
For example, energy consumption could be reduced during sparse occupancy by choosing appropriate thermal set backs, with the PCS providing the additional offset in thermal comfort required for each occupant. 
Our control strategy based on Model Predictive Control (MPC), employs a bi-linear thermal model, and has two time-scales to accommodate the physical constraints that limit certain components of the central HVAC system from frequently changing their set points.

We compare the energy consumption and comfort offered by  our SPOT-aware HVAC 
system with that of a state-of-the-art MPC-based central HVAC system
in multiple settings including different room layouts and partial deployment of 
PCS. Numerical evaluations  show
that our system obtains, in average, 45\% (15\%) savings  in  energy in summer (winter), compared with the benchmark system for the case of homogeneous comfort requirements.  For heterogeneous comfort requirements, we observe 51\% (29\%) improvement in comfort in summer (winter) in addition to significant savings in energy.
\end{abstract}
\end{frontmatter}
%\linenumbers
\section{Introduction}

A typical heating, ventilation, and air conditioning (HVAC) system 
consists of one or more Air Handling Units (AHUs), each with several associated
Variable Air Volume (VAV) units \cite{und02}. 
The AHU chills or heats air to a given set point temperature, 
and the VAV units control the volume of flow of the chilled or heated air 
into a \textit{zone}.
Each zone usually has multiple occupants.
Thus, if these occupants have differing personal comfort requirements,
it may be infeasible to meet them all.

An existing approach to providing individual thermal comfort is to 
deploy a Variable Refrigerant Flow (VRF) system, which can provide fine-grain
thermal control, albeit at a greatly increased capital cost.
Another approach to meeting heterogeneous comfort requirements
(but only in summer) is to use the AHU to
chill air to the lowest desired temperature and provide
a re-heater for each occupant~\cite{und02}.
However, this results in increasing both
the capital cost (for re-heaters) as well as the energy cost, due to wasteful reheating.
Due to these inherent problems,  in most current buildings
the thermal comfort of all occupants is seldom attained
in the presence of heterogeneous
comfort requirements.

%Besides being unable to provide individual thermal comfort,
%traditional HVAC systems are not particularly energy-aware. 
%They employ simple controls such
%as ON-OFF control or Proportional Integral Derivative (PID) control
%which are not energy efficient \cite{naskajsab11}. 
%Hence extensive recent research has proposed advanced control strategies for operation of HVAC systems 
%that reduce the energy consumption without necessarily compromising occupant comfort. The overall goals in this work 
%are to meet individual comfort requirements
%while minimizing energy costs.

To address these issues, in recent work we proposed the
Smart Personalized Office Thermal (SPOT) system \cite{gaokes13,gaokes13a, rabkes16}.
This system combines an off-the-shelf desktop 
%combined 
fan/heater, with 
local temperature sensing and
a computer-controlled actuator to provide individual thermal comfort.
Using a deployment of more than 60 of these desktop
systems over the last two years,
we have found that our personal thermal comfort system can indeed meet
heterogeneous comfort requirements without much additional energy expenditure in a setting where HVAC is not aware of the existence of SPOT.

Given this success,
the following question arises naturally: Assuming widespread deployment of our technology 
(or other similar personal comfort technologies discussed in  Section 6)
how should an existing centralized HVAC system
operate? 
That is, assuming that a
SPOT system is deployed at each occupant's work place and can be used to 
provide personal thermal comfort,
how should we operate the central HVAC system to meet the primary
goal of providing thermal comfort, and secondarily minimizing operational costs?
In this paper, we present one answer to this question, we make HVAC SPOT-aware.

The main benefit of making the central HVAC system SPOT-aware  is that 
overall energy consumption is reduced  during periods of partial occupancy.
When the building is not fully occupied, the central HVAC can provide a base temperature 
 which is lower (resp. higher) than the desired zone temperature in winter (resp. summer).
 The temperature offset can be provided by SPOT's heater for each occupant 
 in the case of heating,  and by its fan, for cooling. 
 Therefore
 energy is not wasted in providing  thermal comfort in unoccupied rooms.
 However when a zone is fully or mostly occupied,  a careful analysis is required to decide whether to change the base
temperature of the HVAC or to use SPOT.
We model the interplay between the central HVAC and SPOT  to analyze the
savings in energy and comfort
that we obtain from the SPOT-aware system. Note that we are not proposing to jointly operate the two systems: indeed, the SPOT system
is unmodified. Instead, the central HVAC system chooses energy-efficient operating
points, knowing that 
limited deficit in user comfort will be made up by SPOT.

Our contributions are as follows. 
\begin{enumerate}
\item 
We present a novel HVAC system  that is composed of several personal thermal comfort systems (called SPOT systems) and a centralized SPOT-aware HVAC. To control it, we present a novel multiple-time-scale controller that combines a two-time-scale MPC-based {\em predictive} controller for the HVAC system (at the 1 hr and 10 minute time scales) using a non-linear thermal model with {\em reactive} control by the SPOT systems at the fastest (30s) time-scale. This formulation assumes that comfort requirements can be met.

\item To compute the personal comfort of an individual, we develop a simplified version of the well-known Predicted Mean Vote (PMV) model that takes air velocity into account and use it as a constraint in our optimization problem.
\item When comfort cannot be met (for example due to heterogeneous comfort requirements), we propose a  modification of the problem to share the discomfort fairly.
In this context, we also propose a new metric to quantify the average discomfort of building occupants.
\item We use extensive numerical simulations to compare our system
 with a central HVAC system which does not use SPOT in both the cases of homogeneous and heterogeneous comfort requirements. We analyze the performance of our proposed system for different building layouts.  We also discuss the pros and cons of having a SPOT-aware HVAC instead of an HVAC which is not aware of the presence of SPOT.  
%We find that our joint system outperforms this benchmark for   We analyse multiple building layouts where we consider zones with  partial deployment of SPOT. 
 \end{enumerate}
 
 The rest of the paper is organized as follows. Section \ref{sec:system} elaborates on the various components in our system and lists our assumptions. A thermal model is derived in Section \ref{sec:sysmodel} and a simplified metric for human thermal comfort is discussed. Section \ref{sec:probform}
explains the principle of our control strategy for operating the HVAC and SPOT systems. Based on this, the optimization problem is formulated where  objective and constraints are listed. 
Finally a method to obtain a solution
for the resulting non-convex optimization problem is briefly discussed. Results are discussed in Section \ref{sec:results}. Related work in literature is provided in Section \ref{sec:related} and the conclusions are in Section \ref{sec:conclusion}.

\section{System and Assumptions}\label{sec:system}
\begin{figure}
\centering
 \includegraphics[scale=0.5]{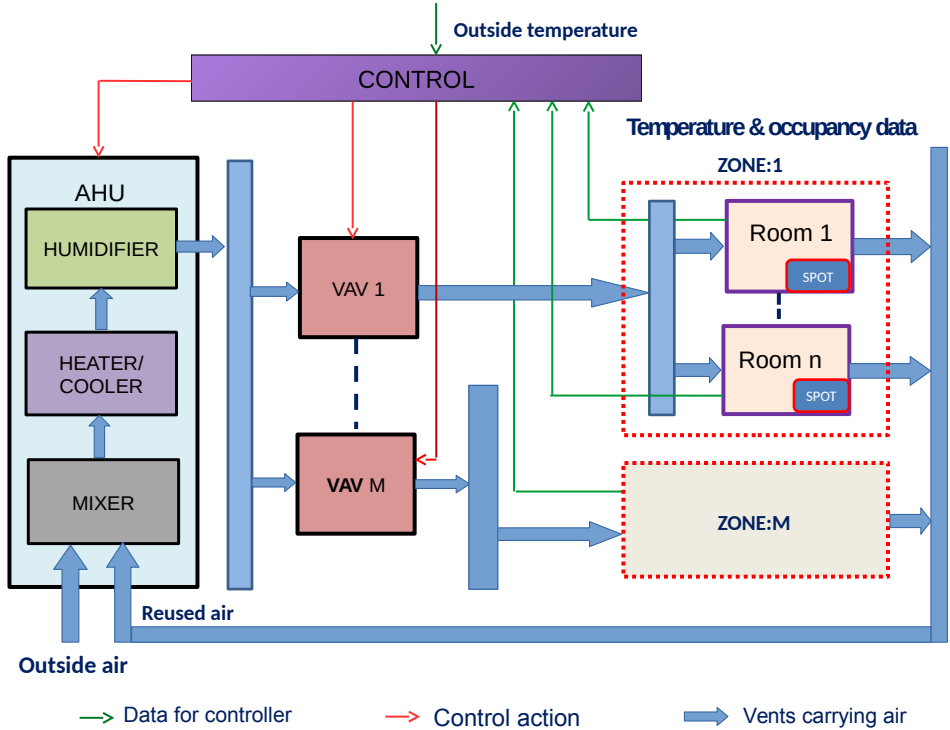}
 \caption{SPOT-aware HVAC system comprising of a centralized controller and multiple SPOT systems}\label{fig:sysmodel}
\end{figure}

In this section, we first describe the system (Figure \ref{fig:sysmodel}), then list our assumptions.
\subsection{The System}

The joint HVAC system consists of the following components:
\begin{itemize}
 \item \textbf{Air Handling Unit (AHU).}
 This unit is comprised of devices such as pumps, heat exchangers, chillers, and boilers
that are used to heat or cool the air obtained from a mixer to a desired temperature,
and a humidifier to control its humidity level. 
The output of the AHU is given to the Variable Air Volume unit, described next.
The AHU air temperature cannot be changed frequently as it can cause damage to the HVAC components \cite{AswMasTan12}. 
Thus, we assume that this value can be changed no faster than once an hour.

\item \textbf{Variable Air Volume (VAV) unit. }
This unit controls
the rate of flow of air from the AHU to the rooms, enabling finer-grained control of room temperature. %\ans{``(on a zone basis)"}\fix{ Mapping of a VAV to a zone is our assumption that is explained later. There is no fixed definition of zone. We call a zone as the rooms that are mapped to the same VAV}
Once the temperature of supply air is set, any further control of 
temperature in the rooms can only be done by varying the rate of flow of air into the rooms\footnote{Note that some VAV units have re-heaters which can further increase the temperature of air. We consider a simple system with no re-heaters.}. 
Unlike the AHU, the VAV's control can be changed fairly often. 
Thus, we assume that this value can be changed every 
10 minutes in our model of the system. 

\item \textbf{Mixer.}
To save energy, instead of only heating or cooling outside air,
some air from the building is recirculated and mixed with outside air in the mixer.
However to maintain air quality, there should always be sufficient fresh air inside the
room. According to ASHRAE (American Society of Heating, Refrigerating and Air-Conditioning Engineers) standards  20 cfm (cubic feet per minute) of fresh air per person
should be supplied.

\item \textbf{The SPOT system.} It is placed at an occupant's work place. Each SPOT has a
heater, a fan (with 10 fan speed settings), a temperature sensor and an occupancy sensor, 
and a controller that \emph{reacts to changes in occupancy
 and comfort level} in the room, by turning on either the heater or fan, within 30 seconds of the change. 
SPOT checks for occupancy and measures temperature every 30 seconds. 
It then computes the personalized PMV\footnote{The PMV index is presented in Section~\ref{subsec:pmv}.}
for the occupant, and if this lies outside the range 
$[-\underbar{$\beta$},\bar{\beta}]$ (a range which is specific to an occupant),  it takes the appropriate control
action. 
If the PMV is larger than $\bar{\beta}$, then the fan turns ON; if it is below -$\underbar{$\beta$}$, the heater turns ON.
%,along with the fan. 
This process is repeated every 30 seconds. 

We have found that SPOT can provide approximately 3 degrees Centigrade of temperature flexibility:
in winter (resp. summer), it can compensate for an AHU temperature set point that is 3 degrees lower (resp. higher) than an occupant's comfort level. Of course, even with SPOT,  the occupant's comfort cannot be guaranteed if it
lies outside this range. 
 \end{itemize}

\subsection{The building}
We study a building with two types of rooms. By a Type $S$ room, we denote
a room with a single occupant with a SPOT system placed at the occupant's work table.
By a Type $\bar{S}$ room, we
refer to a room that does not have a SPOT system. 
%A zone constitutes a set of rooms connected to a given VAV.
In rooms of Type $\bar{S}$, 
comfort is provided by the central HVAC alone;
in rooms of Type $S$, however, 
comfort is provided both
by the central HVAC and SPOT. 
Note that when rooms are unoccupied,
the temperature  is allowed to vary over a wider range, using temperature setbacks,
to reduce energy consumption (details in Section \ref{subsec:optprob}). %\textcolor{red}{[[Rachel, please  verify]]}

\subsection{Modeling assumptions}
 We now enumerate the assumptions made in constructing a mathematical model for the system:
 \begin{enumerate}
 \item We assume that there is a single AHU for the building which supplies air at a chosen supply temperature
and that there is a single VAV unit for each zone that provides a chosen volume of flow of air into the rooms.
In a building with multiple AHUs, each AHU can be separately analyzed using our approach.

\item  For convenience, we assume that all the rooms in a given zone have identical thermal parameters. In practice, this assumption can be easily removed.

\item In Type $S$ rooms, in addition to the temperature and occupancy 
sensors provided by the SPOT system,
we assume that 
there is another temperature sensor in the room which is not in 
close proximity to the SPOT system.
The measurement from this sensor represents the
temperature in the region that is not directly influenced by SPOT. 
We need this sensor to estimate the rate at which heat escapes the 
portion of the room heated by a SPOT system.
\item We assume that in Type $\bar{S}$ rooms there is an occupancy sensor
and a temperature sensor located such that its reading
is representative of the temperature in the entire room. 
\item We assume that the thermal properties of a room can be 
represented using a lumped parameter model.
A Type $\bar{S}$ room is modeled as a single point and  we focus on the temperature at this point.
A Type $S$ room is  modeled with two points: one point represents the occupant's work place with SPOT
and the other point 
 is representative of the part of the room that is not directly influenced by SPOT.
We assume that there is no heat loss in ducts, so that there is no temperature rise or drop between the AHU and the rooms. Again, this assumption is easily removed.
\item We will consider two systems, one comprising an HVAC and no SPOT, called {\bf NS}, and our our system comprising several SPOT systems and a SPOT-aware HVAC, called {\bf SA}. The HVAC in these systems is controlled by a MPC with an horizon of four hours\footnote{We briefly discussed in Section 4 the choice of this time horizon.} and the same forecasts for outside air temperature and occupancy pattern. We assume that these forecasts are available for the entire MPC horizon and are accurate.
We realize that these forecasts, in practice, do have errors, but studying the robustness of
our control to forecast errors is beyond the scope of this paper.

\end{enumerate}
\section{Mathematical Model} \label{sec:sysmodel}
In this section we derive a thermal model for each room type in
the building, then describe the metric that we adopt
to determine the comfort level of an occupant. We start with a model for a Type $\bar{S}$ room,
since it is simpler than that for a Type $S$ room. 
Our notation are in Tables \ref{tab:variables} (for the variables)  and \ref{tab:param} (for the parameters).

\subsection{Thermal model for a Type $\bar{S}$ room}\label{subsec:type2model}

Recall that in a Type $\bar{S}$ room,
there is a single representative temperature for the whole room. 
Based on a first order energy balance,   the continuous time thermal model \cite{KelBor11} of a room $j$ in zone $i$ is as follows. 
% \begin{eqnarray*}
%  \frac{d{T_r}_i}{dt}&=&\frac{1}{C}[-\alpha_o{T_r}_i(t)+\sum_{j=1, j\neq i}^n \alpha_{ij}{T_r}_j(t)]
% - \frac{\rho\sigma}{C} {T_r}_i(t)V(t)  \\
% & &+\frac{\rho\sigma }{C}V(t) T_a(t) +\frac{\alpha_0}{C}T_o(t)+ \frac{1}{C}d_i(t)
% \end{eqnarray*}
\begin{eqnarray}\label{eqn:state:therdyn}
 \dot{x}_{ij}(t)&=&\frac{1}{C_i}[-\alpha^i_ox_{ij}(t)+\sum_{\ell=1, \ell\neq j}^n \alpha^{i}_{\ell j}{x}_{\ell j}(t)]
- \frac{\rho\sigma}{C_i} {x}_{ij}(t)v_i(t)+\frac{\rho\sigma }{C_i}v_i(t)u(t) \nonumber \\
& &+\frac{\alpha^i_o}{C_i}T_o(t)+ \frac{\Oc_{ij}}{C_i}d_{ij}(t) %+  \frac{1}{C_i}Q_hw_{ij}
\end{eqnarray}
where the variables are $x_{ij}(t)$, the temperature of room $j$ of zone $i$, $v_i(t)$, the rate of flow of supply air  into zone $i$, and $u(t)$, the temperature of supply air, all at time $t$.  Note that this model is bi-linear due to the product terms $x_{ij}(t)v_i(t)$
and $v_i(t)u(t)$.
\begin{table}[ht]\caption{Time-dependent variables}\label{tab:variables}
 \begin{tabular}{|l|l|l|}
  \hline
  Notation & Description& Units \\  
  \hline
 $x_{ij}$ & Temperature of room $j$ of zone $i$& $^{\circ}C$\\
$u$ & Temperature of supply air & $^{\circ}C$\\
$v_i$ & Rate of flow of supply air  into zone $i$ & $m^3/s$\\
$w_{ij}$ &Fraction of time the heater of SPOT is ON in one discrete &\\
& time slot in room $j$ of zone $i$ &-\\
$v_{a_{ij}}$ & Speed of the fan in SPOT in room $j$ in zone $i$ & $m/s$\\
$r$ & Ratio of reuse air &-\\
$T_m$ & Temperature of air from mixer & $^{\circ}C$\\
$T_c$ & Temperature of air from cooling unit & $^{\circ}C$\\
\hline
\end{tabular}
\end{table}
%$w$ & Temperature of  air from the cooling %unit &$^{\circ}C$ \\
\begin{table}[ht] \caption{Parameters}\label{tab:param}
 \begin{tabular}{|l|l|l|}
  \hline
 Notation & Description& Units \\   
  \hline
  $n_i$ & Number of rooms in zone $i$ &-\\
$\alpha^i_o$ & Heat transfer coefficient between a room in zone $i$ &  \\
& and outside air & $kJ/K.s$\\
$\alpha^{i}_{\ell j}$ &Heat transfer coefficient between room & \\
& $\ell$ and $j$ in zone $i$& $kJ/K.s$\\
$\rho$ & Density of air & $kg/m^3$\\
$\sigma$ & Specific heat of air  & $kJ/(kg.K)$ \\
$C_i$ & Thermal capacity of a room in zone $i$ & $kJ/K$ \\
$\mathcal{I}$ & Set of zones & - \\

$\Rc_{1}$ & Set of Type $S$ rooms in zone $i$ & - \\
$\Rc_{2}$ & Set of Type $\bar{S}$ rooms in zone $i$ & - \\
%$\tilde{C}$ & Thermal capacity of the occupant & $kJ/K$ \\
$Q_h$ & Heat supplied by SPOT & $kW$\\
$v_a$ & Speed of fan in SPOT & $m/s$\\
$T_o(k)$ & Temperature of outside air at discrete time $k$& $^{\circ}C$ \\
$d_{ij}(k)$ & Heat energy due to internal loads, that is, & kW \\
& lights, equipment and people in room $j$ of zone $i$&\\
& at discrete time $k$ & \\
$\Oc_{ij}(k)$ & Occupancy of room $i$ at discrete time $k$ &-\\

\hline
\end{tabular}
\end{table}

In the following, we discretize time and obtain a discrete model.
This requires two additional assumptions:
\begin{enumerate}
\item We assume that 
the  time
step of the discrete model is $\tau$ (its value will be discussed later).

\item The inputs are assumed to be zero order held with sample rate $\tau$, i.e., they
remain constant during $\tau$.
\end{enumerate}

We employ Euler discretization \cite{kelmabor13} to discretize \eqref{eqn:state:therdyn} 
with a time step $\tau$. 
\begin{eqnarray*}
 \frac{x_{ij}(k+1)-x_{ij}(k)}{\tau}&=\frac{1}{C_i}[-\alpha^i_ox_{ij}(k)+\sum_{\ell=1, \ell\neq j}^n \alpha^{i}_{\ell j}{x}_{\ell j}(k)]
- \frac{\rho\sigma}{C_i} {x}_{ij}(k)v_i(k)+\frac{\rho\sigma }{C_i}v_i(k)u(k)  \\
&+\frac{\alpha^i_o}{C_i}T_o(k)+ \frac{\Oc{ij}}{C_i}d_{ij}(k) %+  \frac{1}{C_i}Q_hw_{ij}
\end{eqnarray*}
The discrete time instant $k$ refers to the time $t_0+k\tau$, where $t_0$ is the start time of the
dynamic process. 
Thus, for all rooms of Type $\bar{S}$ in zone $i$ the thermal dynamics is as follows. 
\begin{eqnarray}
 x_i(k+1)&=&A_{0_i}x_i(k)+A_{1_i}x_i(k)v_i(k) + 
B_iu(k)v_i(k) + D_{1_i}(k)T_o(k) \nonumber\\
&& + D_{2_i}(k)\Oc_{i}(k), %+D_3w_i(k),
\label{eqn:sys:dynamics}
\end{eqnarray}
where $x_i(k),w_i(k)\in\mathbb{R}^n$ are the vectors of all $x_{ij}(k)$'s, $w_{ij}(k)$'s for a given $i$,
\[A_{0_i}=I_n+\frac{\tau}{C_i}\begin{bmatrix}
         -\alpha^i_{o} & \alpha^i_{12} & \dots& \\
          \alpha^i_{21}& -\alpha^i_{o} & \dots & \\
          & & \ddots & \\
          \alpha^i_{1n}& & \dots & -\alpha^i_{o}\\
      \end{bmatrix}
\]
\[A_{1_i}= - \tau \frac{\rho\sigma}{C_i}I_{n_i} , B_i=\tau\frac{\rho\sigma}{C_i}1_{n_i},
D_{1_i}(k)=\tau\frac{\alpha_0}{C_i}1_{n_i}, D_{2_i}(k)=\frac{\tau}{C_i}diag(d_{1i},...,d_{n_ii})\]
\[ \Oc_i(k)=[\Oc_{1i}(k),...,\Oc_{n_i i}(k)]^T,
%D_3(k)=\frac{\tau}{C_i}Q_hI_{n_i},
\]
where $I_{n_i}$ denotes the identity matrix of size $n_i$,
 $1_{n_i}$ refers to a column vector of size $n_i$ with all entries as 1 and $diag(.)$ refers to a diagonal matrix with
 the entries specified.

\subsection{Thermal Model for a Type $S$ Room}\label{subsec:thermalmodel}
In this section, we develop a model for a Type $S$ room. Unlike Type $\bar{S}$ rooms, 
Type $S$ rooms are modeled as two points corresponding to two thermal regions as follows.
\begin{itemize}
 \item Region 1: This region constitutes the occupant's workplace and has the SPOT system. Its thermal level is determined jointly by the central HVAC and SPOT. 
 We denote the temperature of this region in a room $j$ of zone $i$ by $x^1_{ij}$.
 \item Region 2: This region is mostly affected by the HVAC system. SPOT does not influence this region, other than through thermal conduction and convection from the adjacent (SPOT-controlled) zone. We  denote the temperature of this
 region in a room $j$ of zone $i$  by $x^2_{ij}$.  
 \end{itemize}
The thermal levels of the two regions are coupled with each other by conduction and convection. We assume HVAC influences
the thermal level of both regions similarly. We use the thermal model derived in Section \ref{subsec:type2model} to model the
temperature in both regions. Hence, 
the temperature in a room of Type $\bar{S}$ in zone $i$ when SPOT is OFF  is given by the following equation.
\begin{eqnarray}
 x_i(k+1)&=&A_{0_i}x_i(k)+A_{1_i}x_i(k)v_i(k) + 
B_iu(k)v_i(k) + D_{1_i}(k)T_o(k) \nonumber\\
&& + D_{2_i}(k)\Oc_{i}(k),
\end{eqnarray}
We now consider the case where the heater of SPOT is ON followed by the case where the fan of SPOT is ON. Note that this 
is a thermal model for the fastest time scale (i.e., 30 seconds).

When the heater of SPOT is ON, the temperature in both regions will increase. We model this increase in temperature as follows.
First consider region 2, which is directly influenced by  SPOT. Let $Q_h$ denote the power supplied by the SPOT heater and $w_{ij}(k)$ represent the fraction of time in time slot $k$, the SPOT heater is ON.
 Let $\Delta x_{ij}$ denote the 
increase in temperature in region 2 in room $j$ of zone $i$ due to the SPOT heater. We model it by the following equation.
\begin{equation}\label{eqn:deltatemp}
\Delta x_{ij}(k+1) = (1-\frac{\alpha_r\tau}{\tilde{C}_i})\Delta x_{ij}(k)+\frac{\tau w_{ij}(k)}{\tilde{C}_i}Q_{h},
\end{equation}
where  $\tilde{C}_i$ is the thermal capacity of region 2 and
$\alpha_r$ is the heat transfer co-efficient between
the two regions. 
The above equation is written concisely for  all rooms in zone $i$ as follows.
\[\Delta x_{i}(k+1) = \tilde{A}_{0_i}\Delta x_{i}(k)+\tilde{B}_iw_{i}(k),\]
where $\tilde{A}_{0_i}=(1-\frac{\alpha_r\tau}{\tilde{C}_i})I_{n_i}$, $\tilde{B}_i=\frac{\tau}{\tilde{C}_i}Q_{h}I_{n_i}$, and
$w_i(k)=[w_{1i}(k),...,w_{n_i i}(k)]^T$.
In the above equation, we need $\Delta x_{i}(k)$ to compute $\Delta x_{i}(k+1)$. 
This is obtained by taking the difference in the temperatures measured in the two regions at discrete time $k$. 
In the above model only the additional increase in 
temperature caused by SPOT heater to its surrounding is modeled. 
The actual temperature in region 2  is 
$$x^2_{i}(k)=x_{i}(k)+\Delta x_{i}(k).$$
Some heat energy from  SPOT will be 
transferred by  convection to region 1 which is at a lower temperature. Hence the
temperature of region 1 is
\begin{eqnarray}
 x^1_{i}(k)=x_{i}(k) + D_{3_i}\Delta x_i(k-1), 
\end{eqnarray}
where $D_{3_i}=\frac{\alpha_r\tau}{C_i-\tilde{C}_i}I_n$.  

\vspace{6pt}
When SPOT is used in cooling mode, it has no effect on
temperature, but only on the user's perception of comfort. Thus,
if the temperature in both the regions were the same at time $k$, i.e.,
$\Delta x_{ij}(k)=0$, then they continue to remain the same. This is clear from \eqref{eqn:deltatemp}, where we see that when SPOT is in cooling mode, i.e., the heater is OFF, the input $w_{ij}$ is 0.

\subsection{Comfort metric}\label{subsec:pmv}
Human thermal comfort is a function of temperature, as well 
as of humidity, air velocity, clothing level, metabolic rate, and mean radiant
temperature~\cite{fan70}.
For example, in the case of cooling load, a higher air velocity can help the
occupant perceive comfort even when the temperature in the room is higher than a nominal `comfortable' temperature.

A widely-used thermal comfort  metric is the Predicted Mean Vote (PMV) model \cite{fan70} which provides an estimate of the comfort level
based on these parameters.
However computing this metric  is difficult due to its many input
variables and the complex iterative procedure necessary to calculate it. 

Hence we propose a simple analytical  model which is a function of only two variables: air velocity and temperature,
assuming default values for the remaining parameters\footnote{Although humidity can also be controlled by the AHU, for simplicity, we ignore this
in our work.} (different default values for each season).
Specifically,
we assume the mean radiant temperature to be the same as the room temperature (as recommended by ASHRAE 55).
Humidity, clothing level, and metabolic rate  are assumed to be
constant for each  season. 
The typical values for winter and summer for these parameters 
are in Table~\ref{tab:pmv_param}. 
\begin{table}[h]\caption{Parameters for the PMV model for winter and summer}\label{tab:pmv_param}
\center
 \begin{tabular}{|l|l|l|}
  \hline
  Parameter & Winter & Summer\\
  \hline
Humidity $W_r$ & $50\%$ & $50\%$ \\
\hline
Metabolic rate $M$ & 1.1 met & 1.1 met \\
\hline
Clothing insulation factor $I_{cl}$ & 1 clo & 0.5 clo \\
\hline
\end{tabular}
\end{table}

Given these parameters, extracting simplified analytical comfort models 
for the summer and winter seasons involves the following two steps. 
\begin{enumerate}
\setlength{\itemindent}{.5in}
\item [Step~1:] Choosing a functional form for the model.
\item [Step~2:] Obtaining the parameters of the model by fitting the results of the general iterative procedure as explained below.
\end{enumerate}
We tried the following three functional forms where $T$ denotes the temperature and $v_a$ denotes the velocity of air, i.e., the speed of the fan. 
\begin{enumerate}
\item $PMV=f_1+f_2T+f_3v_a $
\item $PMV=f_1+f_2T+f_3v_a^2 $
\item $PMV=f_1+f_2T+f_3v_a^2+f_4v_a $
\end{enumerate}
where $f_1$,$f_2$,$f_3$, and $f_4$ are parameters of the models.  
We obtain two sets of parameters one for each season for each of the above three functional forms.
Note that SPOT could be used in heating or cooling mode
 in both seasons in order to satisfy the  requirements of the occupant.
 
Next we explain our procedure for obtaining the parameters of the above simplified models and to select a model for each season.
We use the online thermal comfort tool from \cite{webberkley} to generate the data required to obtain this 
simplified two-parameter PMV model.
This tool provides the PMV corresponding to the chosen values of room temperature, humidity, air velocity,
clothing level, metabolic rate and mean radiant temperature. 
To get the parameters for winter,  we vary the room temperature from 
$18^{\circ}C$ to $30^{\circ}C$ and the air velocity from $0 m/s$ to $1m/s$ and use the values in
Table \ref{tab:pmv_param} for the remaining parameters and obtain the PMV using the 
online tool.  
Then we 
employ regression to compute the parameters of the simplified model for all the three functional forms along with their Root Mean Square Error (RMSE) values. This is repeated for summer as well. 
We observed that the RMSE value was the smallest for the third functional form.  Hence we use this functional form for our model. 
The  models and the RMSE values are given below.

\vspace{10pt}
\begin{tcolorbox}
{\bf Winter:}  $PMV =  0.25T+0.58v_a^2-1.41v_a-5.47$  \; \; \; \; \; (RMSE: 0.035)\\
{\bf Summer:} $PMV = 0.37T+0.76v_a^2-2.14v_a-9.22$ \; \; \; \;(RMSE: 0.079) 
\end{tcolorbox}

\section{SPOT-Aware HVAC Controller Design} \label{sec:probform}
In this section, we design our SPOT-aware HVAC system. 
%first sketch our control  approach and then formulate our MPC optimization problem.
%\subsection{Controller Sketch}\label{sec:2timescale}
Recall that each SPOT comprises a reactive control mechanism that adapts to
 occupancy and air temperature. 
Our HVAC control strategy aims at computing central HVAC set points knowing that
SPOT reacts autonomously to best meet its owner's preferences.
That is, we \textit{model} how SPOT would react to the central HVAC's set points (using the fast time-scale thermal model in Equation~\ref{eqn:deltatemp}), but
do not control it, letting it operate autonomously.
Instead, we control the AHU, VAV, and reuse parameters at their appropriate time-scales.
Specifically, we change the AHU value every hour, on the hour, and the VAV and reuse values every 10 minutes\footnote{Note that physical constraints on the AHU  can be met as long as the time interval 
between changes is no shorter
than 60 minutes for the AHU. 
Thus, our choice of control times is slightly 
more constrained than strictly necessary. However, this makes the controller design simpler.}.

\subsection{MPC Controller} \label{subsec:optprob}

We now describe a two time-scale MPC-based controller
that runs every $\tau = 10$ minutes (the time step). 
We use the discrete thermal model from Section~\ref{sec:sysmodel},
and assume the availability of 
accurate forecasts of outside temperature and occupancy in each room.
We initially make the simplifying assumption that the MPC controller can meet occupant 
comfort requirements.
We remove this restriction in Section~\ref{infeasible}.

We fix the forecast horizon to be of 4 hours, i.e., 24 time steps\footnote{We found the performance to be almost the same for horizons of 4 and 6 hours.
 Since 6 hours increases the computational burden with little gain in performance, we used 4 hours for the analysis in our paper.}. 
 At the beginning of each time step, we obtain (or revise) the forecasts for room occupancy and outside temperature for the entire horizon and re-compute all controlled
 values: ratio of reuse air, volume of air flow into each zone, as well
 as the predicted SPOT status in each room with SPOT, i.e., if SPOT is ON or OFF and its action if it is ON.
 We also update the value of  supply air temperature once every hour.

The MPC objective is to minimize the total energy consumption subject to the constraint that each occupant comfort is always met,
that is, if an occupant is thought to be present, the PMV level in the room is guaranteed to be in his or her desirable range. 

Let $\I$ denote the set of zones. 
The power consumed by the heating process, $P_h(k)$ and the cooling process, $P_c(k)$ are determined based on the air-side
thermal power as follows.
 $$P_h(k)=\sum_{i \in \I} v_i(k)\theta_1(u(k)-T_c(t)),$$
 $$P_c(k)=\sum_{i\in \I} v_i(k)\theta_2(T_m(k)-T_c(k)),$$
 where 
 $\theta_1=\frac{\rho\sigma}{\eta_h}$, $\eta_h$ is the efficiency of the heating unit;
 $\theta_2=\frac{\rho\sigma}{\eta_c}$, $\eta_c$ is the efficiency of the cooling unit; $T_c$ is the temperature  of the air coming from the cooling unit; and $T_m$ is the temperature of 
 the air coming from the mixing unit. Another component that consumes energy is the fan that
blows the supply air. The power consumed by the fan in HVAC is given by the following model from \cite{KelBor11}.
$$P_f(k) =\theta_3 (\sum_{i\in \I}v_i(k))^2$$
where  the value of $\theta_3$ is given in Table \ref{table:parametervalue}. 
Let  $\Rc_{1}$
denote the set of rooms with SPOT in zone $i$.
%\green{Why does i appear twice in the notation for $\Rc$?}.
The heater and fan of SPOT also consume power  which are given by 
\[Pspot_h(k)=\theta_4 \sum_{\substack{i\in \I,\\ j\in\Rc_{1}}}w_{ij}(k),\,\, Pspot_f(k)=\theta_5 \sum_{\substack{i\in \I,\\j\in\Rc_{1}}}v_{a_{ij}}(k),\]
where  the values of $\theta_4$ and $\theta_5$ are given in Table~\ref{table:parametervalue}.
Hence the energy consumption $J$ which we aim to minimize is given, for the time horizon $N$, by 
$$J=\sum_{k=0}^{N}[P_h(k)+P_c(k) +P_f(k)+Pspot_h(k)+Pspot_f(k)]\times \tau$$
Note that, at a given point of time either the heating unit or the cooling unit is employed. This implies that  in the objective function, at any time $k$, both $P_h(k)$ and $P_c(k)$ cannot be non-zero (but they can both be zero).

\vspace{10pt}
\noindent The MPC is subject to the following constraints:

\begin{enumerate}
\item \textbf{Comfort requirements}: 
For a Type $\bar{S}$ room, we use the single temperature measurement available from the sensor in that room. The constraint is as follows:
\[\underbar{$\kappa$}\leq x_{ij}(k) \leq \bar{\kappa},\forall i\in\I,j\in\Rc_2\]

For Type $S$ rooms where personalized comfort is provided using  SPOT, PMV is used as a metric for comfort. 
We use the model obtained in Section \ref{subsec:pmv} to compute the PMV in each room.  Let $\Pm_{ij}$  denote the 
PMV in  room $j$ of zone $i$. 
We use the temperature of region 2 (SPOT region), where the occupant is present, to compute $\Pm_{ij}$.  
Let $\underbar{$\beta$}_{ij}$ and $\bar{\beta}_{ij}$ denote the preferred lower and upper limits for PMV
in Type $S$ room $j$ in zone $i$ when it is occupied. Then, the constraints are:
\[  \underbar{$\beta$}_{ij}\leq \Pm_{ij}(k) \leq \bar{\beta}_{ij},\forall i\in\I,j\in\Rc_1.\]
Though the occupant's comfort level is not determined by the thermal level of region 1, 
we have some constraints on the temperature of this region. This ensures that this region has some acceptable thermal level if not the strict requirements of region 2.
%\textcolor{red}{[[why? has this been explained? I don't think so. By the way do not erase my comments, comment them, i.e., put a $\%$ before them when you have taken care of them.]]}
\[\underbar{$\gamma$}\leq x^1_{ij}(k) \leq \bar{\gamma},\forall i\in\I,j\in\Rc_1\]

If a room is not occupied (irrespective of its type), there is a constraint to ensure that the building temperature remains at an acceptable level in case of sensor failure. Also, for places without occupancy sensors such as corridors we need a minimum temperature in winter or a maximum temperature in summer.
\[  \underbar{$\gamma$}\leq x_{ij}(k) \leq \bar{\gamma},\forall i\in\I,j\in\Rc_1\cup\Rc_2\]
%Also the mechanical limitations of various components do
%not permit sudden changes and overshoots in the variables. Hence the building is maintained at an appropriate thermal level even if unoccupied.
\item \textbf{Input constraints}: There are certain constraints on inputs due to  the ratings
of actuators and  physical limitations of the components of HVAC.  
The limitation of the  heating unit determines the  maximum limit of the supply air temperature. Similarly the limitation of the cooling unit  decides on the minimum temperature of the supply air.  In the case of the air flow, a minimum 
 rate has to be  maintained in each zone so as to have adequate amount of outside air when occupied. The fan capacity and size of vents determine the upper bound on $v$. % \textcolor{red}{[[This does not explained the upper bound on v.]]}
\[\underbar{$u$} \leq u(k) \leq \bar{u}, \forall k\]
\[\underbar{$v_i$} \leq v_i(k) \leq \bar{v_i}, \forall k, i\in \I\]
When the room is unoccupied, the lower limit of $v_i$, i.e., $\underbar{$v$}_i$ is set to 0. 
%This is because there is no need for any 
%ventilation when the room is unoccupied \textcolor{red}{[[I don't think this is true]]}. \fix{this is our assumption. there is no need of fresh air to be  supplied if not occupied. However, if the thermal level drops below the required level (even if unoccupied)  then the air flows in to bring up the thermal level. Only the lower limit is set to 0 and not that $v_i$ is forced to be 0 when unoccupied}\ans{This is a bad assumption. I think we should remove it.}\fix{Our results are computed based on this assumption for both our system and the benchmark system. Maybe the savings would be the same if we relax this assumption. But we cant be sure. In another paper also they have kept the lower limit to be 0.}
\item \textbf{$u$  can be changed only once in an hour}, i.e., $u$ is
constrained to be the same for all time steps within an hour. We explain the mathematical formulation
of this constraint for $u$. We denote the discrete time $k$ (i.e., the $k$th slot of 10 minutes) in a day using the pair $(p,q)$, where $k=6p+q$, $p\in \{0,1,...,23\}$ and $q \in \{0,1,...,5\}$. Hence at each time, $(p,q)$
we compute the optimal values for our controlled parameters using MPC. Since our horizon is 4 hours, at each discrete time, we compute 24 values of $u$. Consider the following two cases:
\begin{itemize}
\item [Case (i):] The time $(p,q)$ is 
%\textcolor{red}{[[I don't understand what you mean by this. The MPC is recomputed every time slot which you should say. DO you mean if the time slot is the beginning of an hour this is case (i) and otherwise it is case (ii). If it is this please say it that way. You cannot erase my comment if you do not answer it. The sentence means nothing]]} 
at the beginning of an hour in a day, i.e., $q=0$. 
Then for the 24 values of $u$ to be computed, we have 4 constraints (one per each hour in the time horizon)
%\[u(6z+1)=u(6z+2)=...=u(6z+6), \mbox{ for } z=0,1,2,3.\]
\[u(6z)=u(6z+1)=...=u(6z+5), \mbox{ for } z=p,p+1,p+2,p+3.\]
%\textcolor{red}{[[Again I don't understand, z is not defined and it takes 4 values, so you have 4 constraints and not 24. Also I don't see p in the constraints]]}
%\fix{z is just a variable no specific meaning. Yes there are 24 variables but only 4 constraint equations. p denotes the hour. The constraints are same every hour. Only beginning of an hour or in between an hour matters. So only q comes in the constraints.}
\item [Case (ii):] The time $(p,q)$ is not at the beginning of an hour, i.e., 
%\textcolor{red}{[[same remark, I think that you mean "computed"]]} 
$q \neq 0$. Then the constraints on $u$'s are as follows.
\[u(6p+q)=u(6p+q+1)=...=u(6p+5),\]
\[u(6z)=...=u(6z+5), \mbox{ for } z=p+1,p+2,p+3.\]
\[u(6p+23-q)=...=u(6p+23).\]
%\[u(1)=u(2)=...=u(6-q),\]
%\[u(6z-q+1)=...=u(6z-q+6), \mbox{ for } z=1,2,3.\]
%\[u(6-q+19)=...=u(24).\]
%\textcolor{red}{[[Howcome there is no p?]]}
%\fix{same answer as for previous comment} 
Note in this case, $u(6p+q)$ need not be computed. It takes the value that was computed from the previous instance of the MPC that is currently being implemented as the set point in AHU.
The constraint equations are determined based on the value of $q$. 
\end{itemize}The constraints for both the cases are written in a concise form in Table~\ref{table:optprob}
% \textcolor{red}{[[I don't understand why there is no p. I also don't like the fact that we take almost a full page to describe simple notations....]]} 

\item \textbf{We reuse the exhaust air from the rooms}. The temperature of the exhaust air, $T_e$  is assumed 
to be the average temperature of all rooms. Let $r$ be the ratio of exhaust air to the total
air taken to the AHU from mixer unit. Hence the temperature of air coming out of the mixer unit, $T_{m}$ is given by
\begin{equation}\label{eqn:reuse}T_{m}(k)=r(k)T_e(k)+(1-r)T_o(k).\end{equation}
\[0\leq r \leq \bar{r}\]
Here $r=0$ means no reuse of exhaust air (economizer operation). The upper limit, $\bar{r}$ is determined by the amount of outside fresh air
required for maintaining good indoor air quality.
%\textcolor{red}{[[say why]]}. 
\item  \textbf{The heater can only increase the temperature and the cooler can only decrease the temperature}.
\[ T_c\leq T_m, \,\, u\geq T_m \] %\textcolor{red}{[[I don't understand why these 2 constraints are on $T_c$]]} 
%\fix{We assume first there is mixer unit and then cooling unit and then heating unit. At a time only one unit will be employed, i.e. either we heat the air or cool the air. But I do agree this is not clear. So I have modified both in terms of Tm (mixing unit). It is same as the previous constraints but this is more clear I believe.}\ans{Keshav, please check}
\item \textbf{SPOT is ON only if the room is occupied.} 
 \begin{equation*}w_{ij}\leq \Oc_{ij}, v_{a_{ij}}\leq \Oc_{ij} \end{equation*}
 \begin{equation*} 
 v_{a_{ij}}\in \mathcal{V} = \{0,0.1,0.2,\dots, 1\}
 \end{equation*}

%\subsection{Relaxing the Integer Constraint}
Note that the variable $v_{a_{ij}}$ is constrained to belong to the discrete set $\mathcal{V}$. We propose to relax this integer constraint as follows:
\[0\leq v_{a_{ij}} \leq \Oc_{ij}\bar{V}_a,\,\, \bar{V}_a=1~m/s\]
%Recall that there are 10 distinct speed levels, %\textcolor{red}{[[I think that it is the first time that you talk about this. It should have been mentioned in the system description.]]\fix{It is mentioned in the system description}}, 
Then, we  use the value in $\mathcal{V}$ which is closest to the computed value of  $v_{a_{ij}}$.\end{enumerate}

%\textcolor{red}{[[I don't think the previous paragraph is correct. $v_{a_{ij}}$ is defined in Table 1 as the speed of the fan in m/s. So what you can say is that $v_{a_{ij}} = n_{ij}^s S_s$ where $S_s$ is the $s^{th}$ speed of the fan and $n_{ij}^s \in \{0,1\}$ with $\sum_{s=0}^{10} n_{ij}^s = 1$. Then you can relax the integer constraint on $n_{ij}^s$.]]}\fix{If am not wrong, here a convex combination of fan speeds is taken but that would not result in an element in the constrained set as the set is not convex to begin with. We still need some rounding off to do. I had made a mistake in the set to which $v_{a_{ij}}$ belongs in the previous formulation. I have corrected that. Hope the present content is fine.}

%\textcolor{red}{[[With respect to $w_{ij}$ I think that there is NO reasons to introduce it as a binary variable and then transform it immediately into a real value KNOWING that the real value model is better to start with. We should define it immediately as the proportion of the time slot SPOT is on. So please rewrite all this]]} 
%\textcolor{red}{[[what is the purpose of this last sentence. Comes from nowhere....]]} 
Note that the  formulation so far is for a system with a single VAV. The optimization problem is generalized for a building with multiple VAVs in Table \ref{table:optprob}.

The optimization problem formulated above (and summarized in Table~\ref{table:optprob}) is non-convex due to the bi-linear thermal model. We discuss next how to deal with this complexity.

\subsection{Solving The Non-convex MPC}

It has been suggested in \cite{Bie98} that 
Sequential Quadratic Programming (SQP) techniques can be used to handle non-linear
MPCs. Solvers like SNOPT and NPSOL use an SQP algorithm to 
compute a solution though there is no guarantees that the solution is optimal. In the following, we use SNOPT to solve our optimization problem, though we do so with some care. 
Since the problem is non-convex the solution provided by 
SNOPT depends on the initial guess provided to the solver. 
To avoid the pitfall of a local solution, we investigated if
different initial guesses yielded widely different solutions. We performed the analysis for 1000 randomly generated initial guesses, with a uniform distribution,
within the specified ranges for the input variables, for several instances of the optimization problem. 
We concluded that for each instance of the optimization problem, if we compute the solution for 15 randomly generated initial guesses and take the minimum value among these as the solution, then 
we almost always find  a solution that is within 5\% of the best value obtained for the 1000 guesses.
In short, we observed that by using 15 random initial guesses we could avoid 
the risk of a bad local optimum obtained by using just a single initial guess or the default initial guess of the solver.

% \subsection{Reactive Control Using SPOT}
% Note that \textcolor{red}{the MPC described above is working on two time-scales (10 minutes and 1 hour [[Right?]]). SPOT will react autonomously  within 30 seconds to a change in occupancy status and can provide approximately 
% 3 degrees of flexibility in temperature setback.  [[I am confused by the rest of this subsection. What are we trying to say? This should be a comment not a subsection]]} 
% Knowing this, the HVAC central controller uses the MPC to compute the optimal AHU and VAV set points, 
% modeling the additional gain in comfort from
% the existence of SPOT in certain rooms. 
% More specifically, the MPC controller estimates the gains in user comfort due to SPOT's heater and fan
% (using the models described in  Section~\ref{subsec:thermalmodel} and Section~\ref{subsec:pmv})
% to choose the optimal AHU and VAV setpoints, thus allowing setbacks, when feasible.
%Hence, even though SPOT  operates by reacting to occupancy, the performance expected by the central controller is achieved. 
%
\subsection{When Comfort Requirements Cannot Be Met}\label{infeasible}
Thus far, we have assumed that the HVAC system can meet  occupant comfort requirements. 
We now consider the more realistic case when occupant comfort requirements may not necessarily
be met. This can be due to heterogeneous comfort requirements, or even possibly due to the homogeneous requirements
falling outside the range that can be provided by the HVAC system. 

When comfort requirements cannot be met, we re-formulate
the earlier objective function  (to minimize energy consumption)
to include the additional goal of minimizing discomfort.
This is done by adding a penalty term for discomfort, 
while simultaneously relaxing the comfort constraints,
as described next. 
Recall that $P_{ij}$ denotes the PMV in room $j$ in zone $i$ 
and [$\underbar{$\beta$}_{ij}$, $\bar{\beta}_{ij}$] is the range
of acceptable  comfort in that room when occupied.  We relax the 
comfort requirements for all rooms indexed by $j$ and zones indexed by $i$ as follows:
 \begin{align*}\label{eqn:new_comfort}
\underline{\beta}_{ij} - \epsilon_\ell(ij) \leq P_{ij} &\leq \bar{\beta}_{ij} + \epsilon_h(ij)  ,\\ 
% P_{ij} &\geq \underline{\beta}_{ij} - \epsilon_\ell(ij),\\
 \epsilon_\ell(ij) \geq 0,  \; \; \epsilon_h(ij) &\geq 0 
 \end{align*}
\noindent where  $\epsilon_\ell(ij)$  and $\epsilon_h(ij)$ are  variables that depict the
 deviations from the upper  and lower limits  respectively of the comfort requirements.
We penalize the deviations in comfort by adding the term in Eq.~(\ref{eqn:addnterm}) to the objective function. 

%\fix{[[If it is an additional term, you should not put the minimize  since it seems that you are ONLY minimizing this term, I removed it. ALSO $\epsilon_l$ AND $\epsilon_h$ HAVE NEVER BEEN DEFINED. please explain why this is a good way to take comfort into account, why the sum? why the product?]]}.
  \begin{equation}W \prod_{\substack{i\in\I,\\j\in\Rc_1}} (1+\epsilon_\ell(ij) +\epsilon_h(ij)),\label{eqn:addnterm}\end{equation}
where, $W$ is a weighing factor which determines the comfort energy trade-off and this is set to a high value (see Table \ref{table:parametervalue}) to ensure minimum discomfort. 
This additional term is the product of comfort deviations
 in each room to fairly distribute discomfort amongst the rooms.
%The performance of the system for various values of $W$ requires a more elaborate analysis and is not pursued in this paper.
%
\section{Results and Discussion} \label{sec:results}
In this section we evaluate
the performance, in terms of energy and comfort, 
of our proposed SPOT-aware HVAC system (denoted \textbf{SA}).
A comprehensive recent survey of HVAC control techniques concludes that
``Compared with most of the other control techniques, MPC generally provides superior performance in terms 
of lower energy consumption, better transient response, robustness to disturbances, and consistent performance under varying conditions"~\cite{mpcsurvey}. For this reason, we compare SA only with a conventional MPC-based HVAC system that is similar in spirit to SA but does not have SPOT deployed (denoted \textbf{NS}). NS does not model the state evolution of the deployed SPOT instances, but otherwise
uses the same controls as SA.

%[SU:] A system that is unaware of SPOT deployments but is otherwise identical to SA. The principal difference between the SU and SA controllers is that the SU controller does not model the state evolution of the deployed SPOT instances.

The values of the parameters used in our  numerical studies are listed in Table \ref{table:parametervalue}. 

\subsection{Expected Performance Outcomes }
We first discuss our expectation of  the relative performance of the two systems.
To do so,
recall that the deployment of SPOT systems has two distinct and separable benefits:

\begin{adjustwidth}{0.5in}{}
\begin{description}
\item  [Benefit 1] It provides a few degrees worth of heating and cooling, at an energy cost that we expect to be lower than with a centralized HVAC system, since it does need to heat/cool  an entire zone, or even an entire room, only the space around the occupant.
\item [Benefit 2] It allows heterogeneous comfort requirements to be met, at an additional energy cost 
\end{description}
\end{adjustwidth}

In the following, we quantify these benefits. first study homogeneous comfort requirements, then 
heterogeneous comfort requirements for two types of scenarios: one
where there is a SPOT system in every room,
and one where there are rooms without SPOT. 

%As a preview of our results, 
%we find that the NS system is unable to meet occupancy
%comfort requirements. We also find that the reduction in energy using SA due to partial
%occupancy more than compensates for the increase in energy cost to meet heterogeneous
%comfort requirements.
\subsection{Homogeneous Comfort Requirements}

We compare
the performance of the SA and NS systems, in terms of occupant comfort and energy use,
with full and partial SPOT deployment. This is because buildings have common areas (as opposed to 
private workspaces)
that are not suitable for deployment of the SPOT system. We would also like to compare their performance both 
in the simple case of a building with a single zone, and in the more complex case of 
a building with multiple zones. 
Accordingly, we define the following three scenarios:

\begin{adjustwidth}{0.5in}{}
\begin{description}

 \item [Scenario 1 (S1)] (1 zone, full SPOT deployment): This building has five identical Type $S$ rooms, that is, with a SPOT in each room. The rooms are identical, adjacent, and are thermally insulated from each other.
 \item [Scenario 2 (S2)] (1 zone, partial SPOT deployment): This building is the same as in S1, but it has four rooms with SPOT (Type $S$) and one room does not have SPOT (Type $\bar{S}$). Comparing the performance of SA in Scenario 2  and Scenario 1 allows us to determine whether SA performs well when SPOT is not deployed in every room in the building.
 \item [Scenario 3 (S3)] (2 zones: one with full and one with partial SPOT deployment): 
In zone 1 of this building, there are five identical Type $S$ rooms (with SPOT in each room). 
In zone 2 there is one large Type $\bar{S}$ room (without SPOT) corresponding to a meeting room/class room.
As before, the rooms are all thermally insulated from each other.
\end{description}
\end{adjustwidth}
\begin{figure}
\begin{minipage}{0.5\textwidth}
\centering
\includegraphics[width=0.7\textwidth]{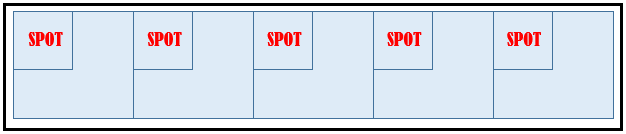}
\captionsetup{labelformat=empty}
\caption{Scenario 1}
\centering
\includegraphics[width=0.7\textwidth]{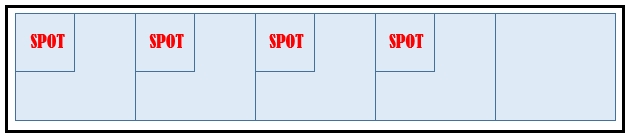}
\caption{Scenario 2}
\end{minipage}
\begin{minipage}{0.5\textwidth}
\centering
\includegraphics[width=0.7\textwidth]{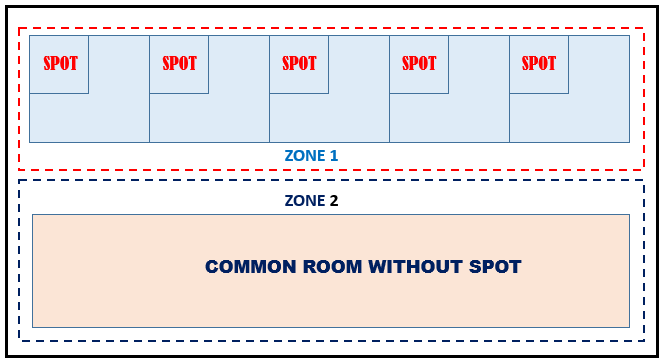}
\captionsetup{labelformat=empty}
\caption{Scenario 3}
\end{minipage}
\caption{Layout for the three scenarios}
\end{figure}

\begin{center}
\begin{figure}[!ht]
     \subfloat[Outside temperature \label{subfig-1:dummy}]{%
       \includegraphics[width=6cm, height=4cm]{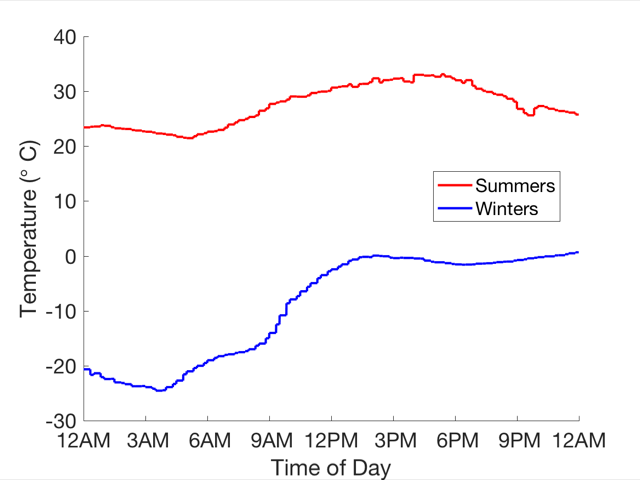}
     }
     \subfloat[Occupancy pattern for the 5 rooms in a zone\label{subfig-2:dummy}]{%
       \includegraphics[width=9cm, height=4cm]{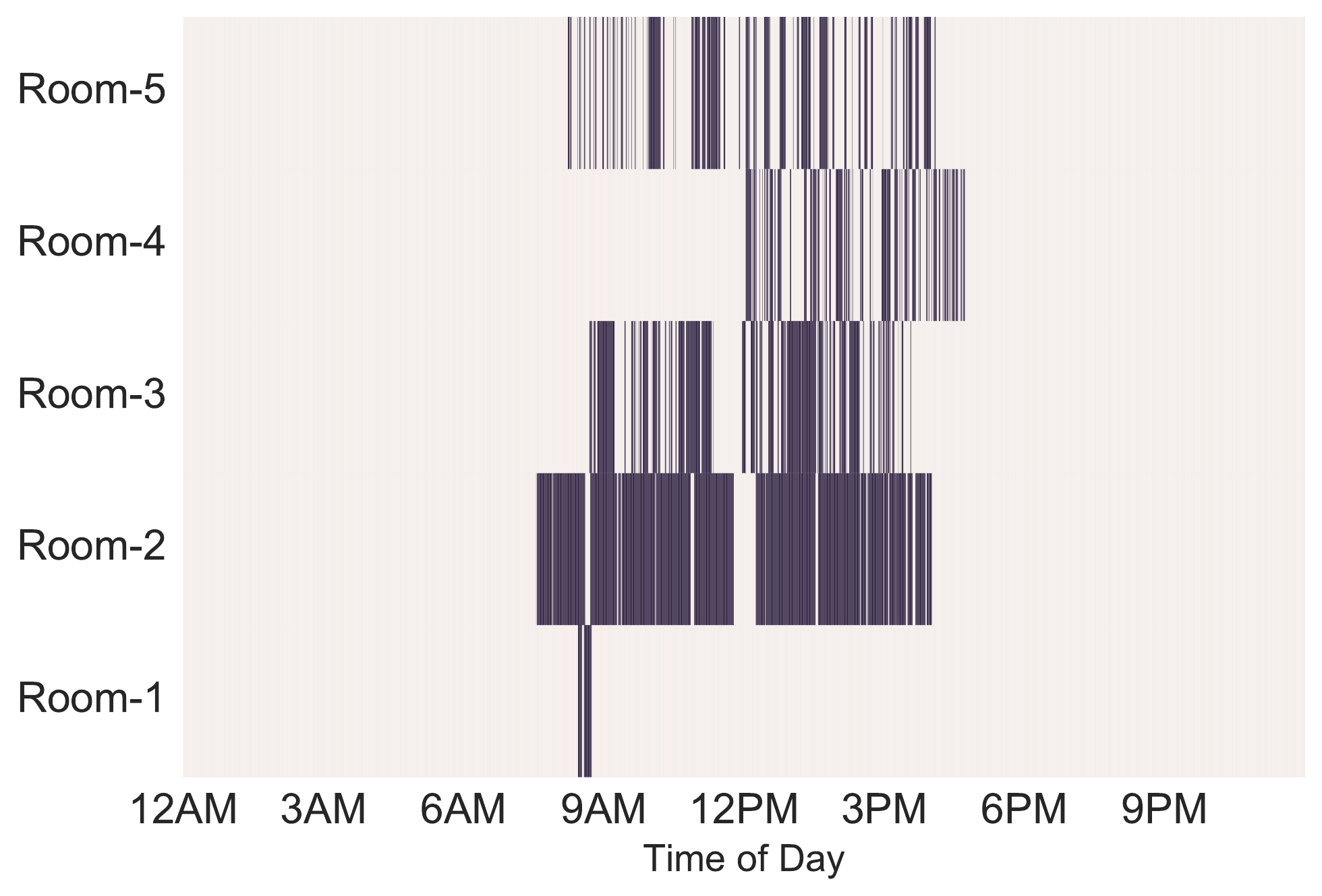}
     }
%     \hfill
     \caption{Outside temperatures and occupancy patterns for a day}
     \label{fig:temp_ocp}
\end{figure}
\end{center}

%\red{Please add a figure showing the layout.}
For all three scenarios, we used realistic room occupancy data obtained by placing 
passive infrared sensors on occupants' desktops as part of
the SPOT* project~\cite{ocpdata}. In total,
we collected more than 300,000 hours of occupancy data collected at
30s intervals from
about 60 offices over a period of approximately 1 year. We also used actual temperature data
from University of Waterloo's weather station \cite{tempdata}. 
For summer we chose days from the months of June, July and August and for winter we 
chose days from the months of December, January and February. Figure \ref{fig:temp_ocp} shows the outside temperature for a summer and a winter day and sample occupancy patterns in a zone.
%\textcolor{green}{
%Show one figure for typical occupancy for one room for 5 rooms for one day. Show temp of typical sumer and winter day (on the same graph). Put the rest in github and put a link in the paper.}

The comfort requirements, which are based on ASHRAE standards, are given in Table \ref{tab:pmvvalues:hom} in Appendix.
These values correspond to We computed numerical results using Matlab and employed 
the standard SNOPT solver
for solving the optimization problem. To compute
thermal evolution in the building,
we designed and implemented 
a custom building simulator in C++~\cite{ThermalSim}.
We did not use a standard building simulator, such as the ones
described in Reference~\cite{hong2000building}, because it was
challenging to incorporate personal thermal comfort systems into them.
In contrast, our simulator implements the discretized thermal 
models for Type $S$ and Type $\bar{S}$ rooms described in Sections 3.1 and 3.2 respectively,
which we found to be a straightforward task.

We found that, in all the three scenarios, 
both systems (NS and SA)  were able to meet the homogeneous comfort requirements that we considered. Hence we do not analyze the performance with respect to comfort for this case.  Instead, we 
%measure the supply air temperature and 
compute the energy consumption for 50 days in summer and for 50 days in winter for the two systems in the three scenarios.

\subsubsection {Energy Consumption}
\begin{figure}[!ht]
     \subfloat[Scenario 1\label{subfig-11:dummy}]{%
       \includegraphics[width=0.33\textwidth]{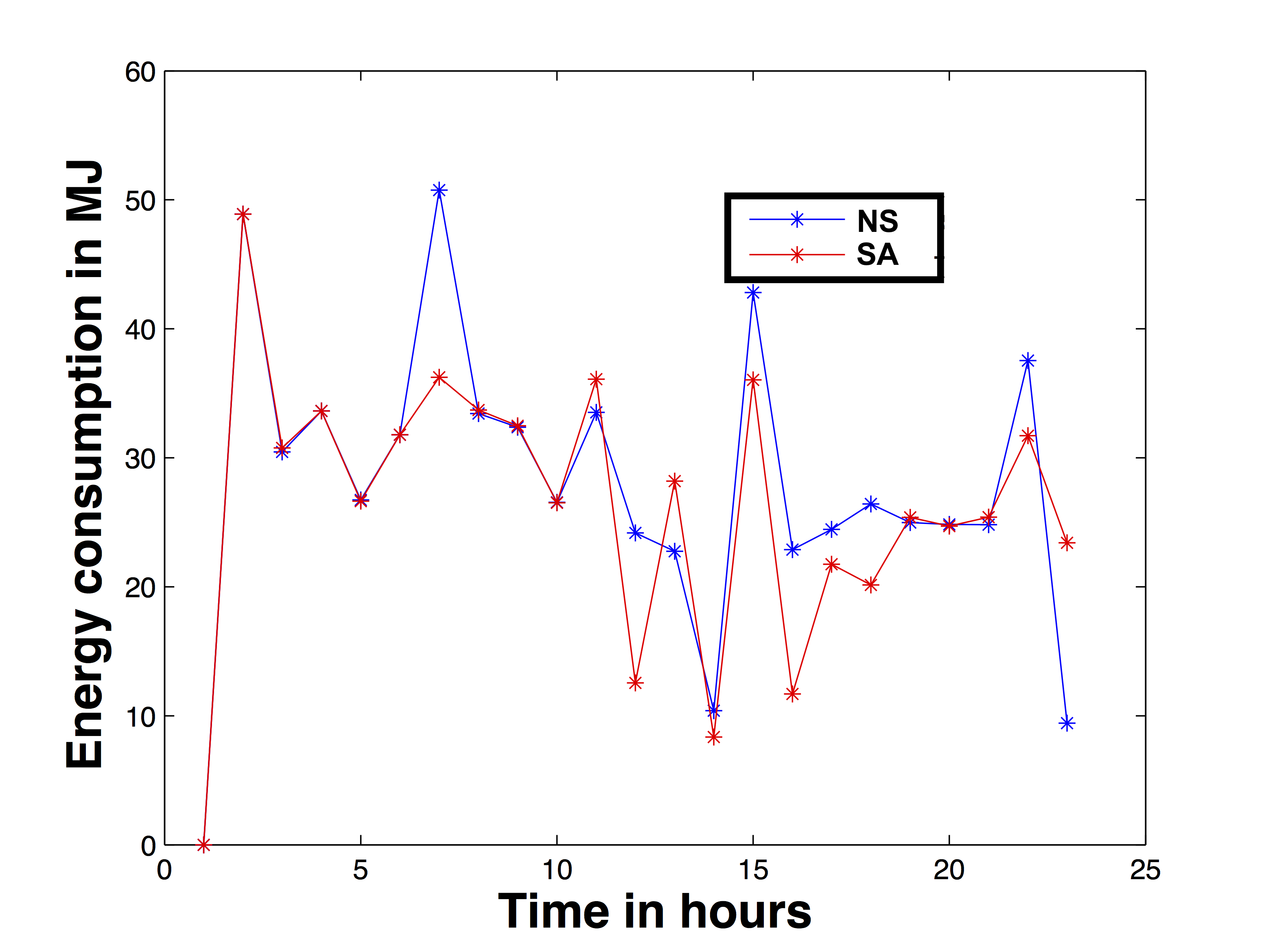}
     }
     \subfloat[Scenario 2\label{subfig-21:dummy}]{%
       \includegraphics[width=0.33\textwidth]{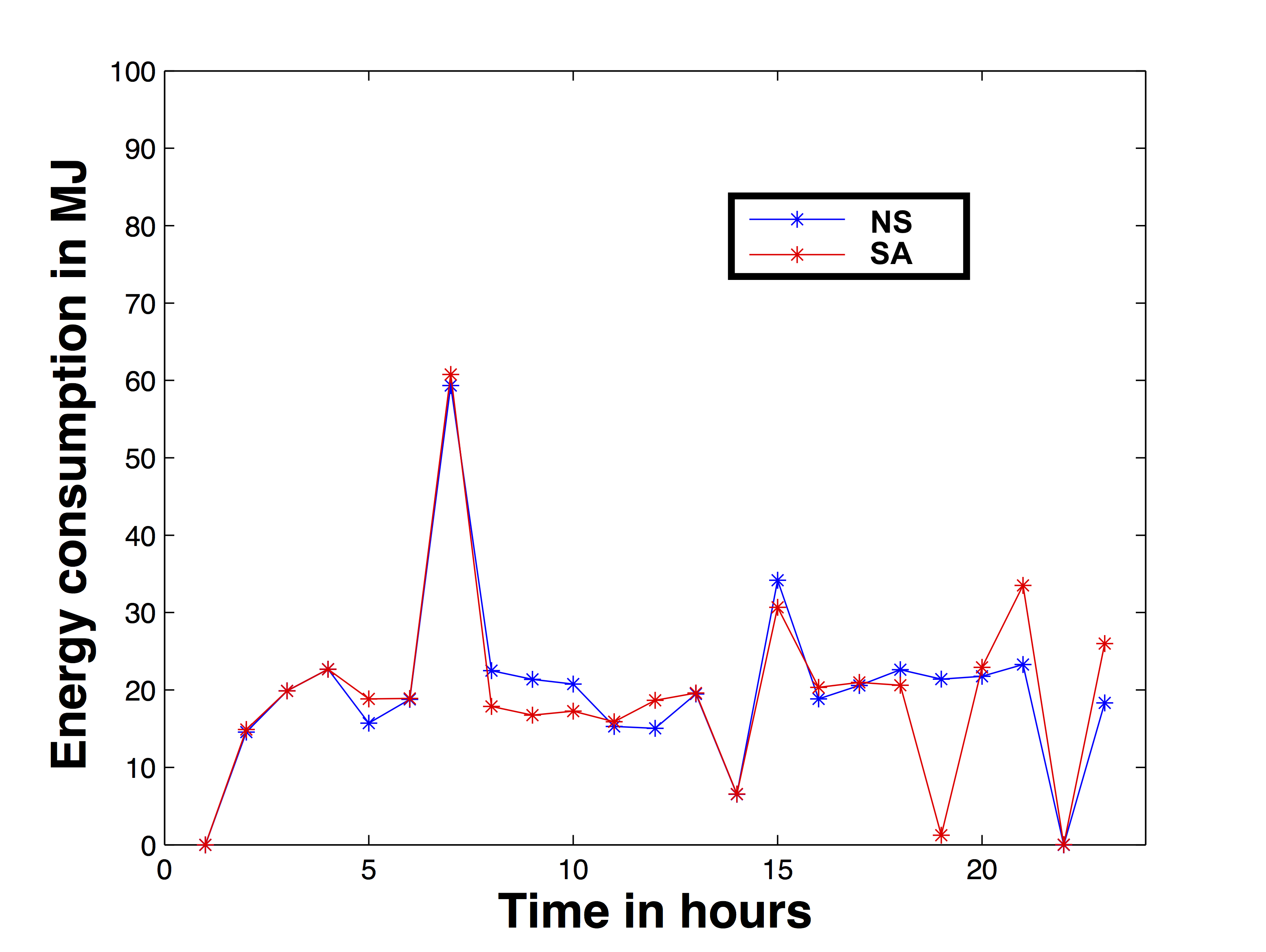}
     }
%     \hfill
     \subfloat[Scenario 3\label{subfig-31:dummy}]{%
       \includegraphics[width=0.33\textwidth]{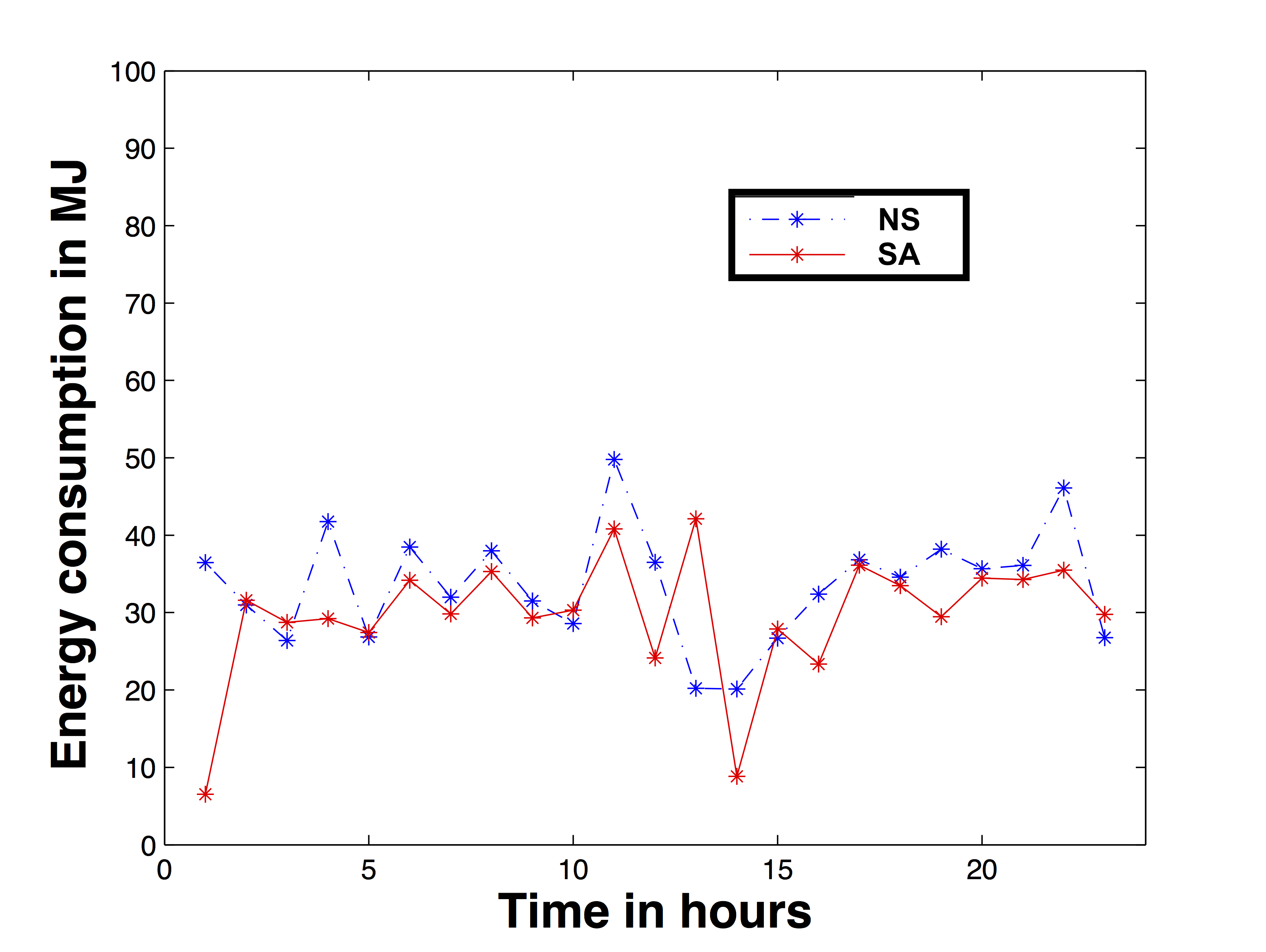}
     }
     \caption{Comparison of energy use by SA and NS for one typical \textbf{winter} day.}
     \label{fig:energy_winter}
\end{figure}

\begin{figure}[!ht]
     \subfloat[Scenario 1\label{subfig-12:dummy}]{%
       \includegraphics[width=0.33\textwidth]{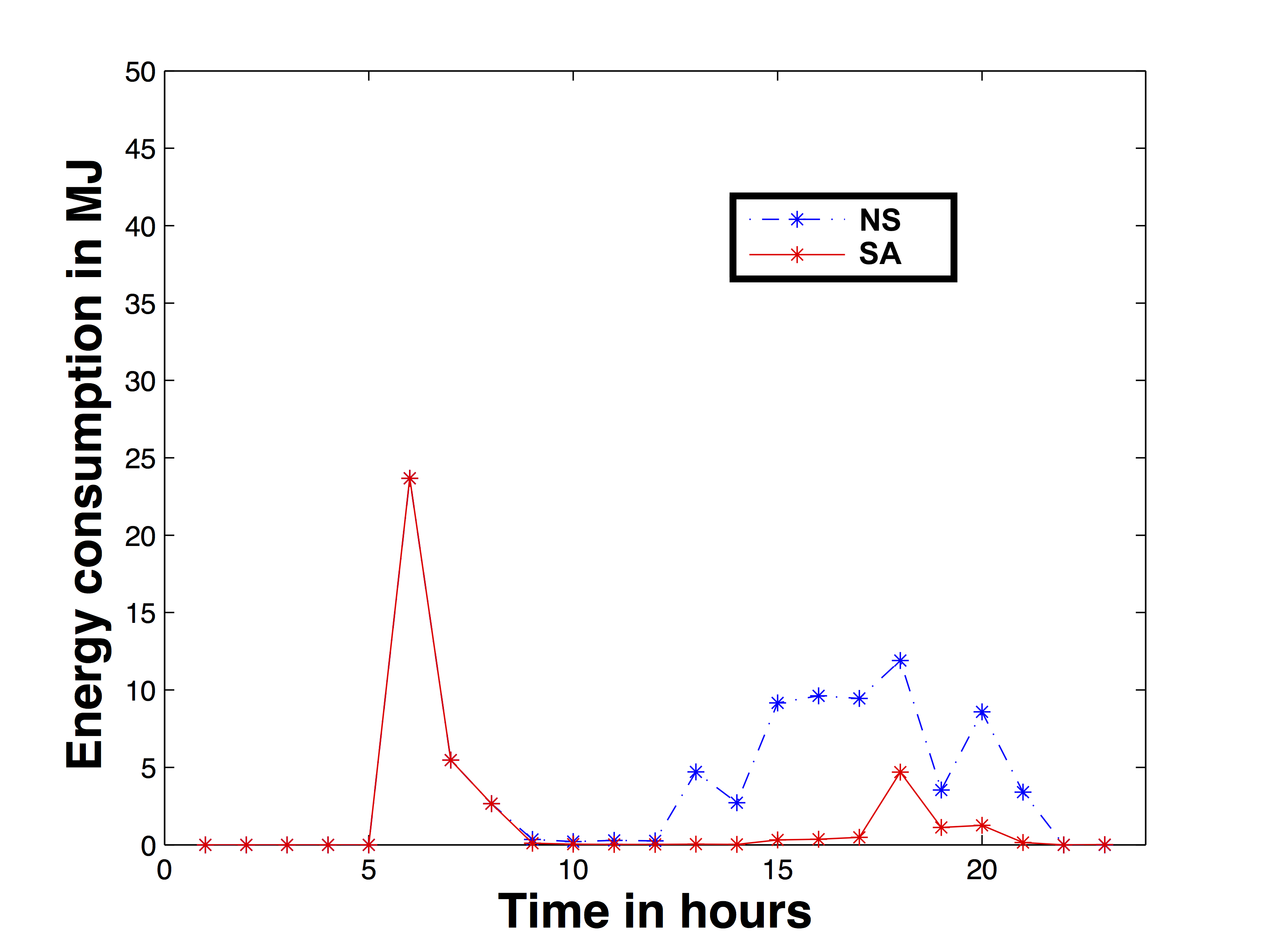}
     }
     \subfloat[Scenario 2\label{subfig-22:dummy}]{%
       \includegraphics[width=0.33\textwidth]{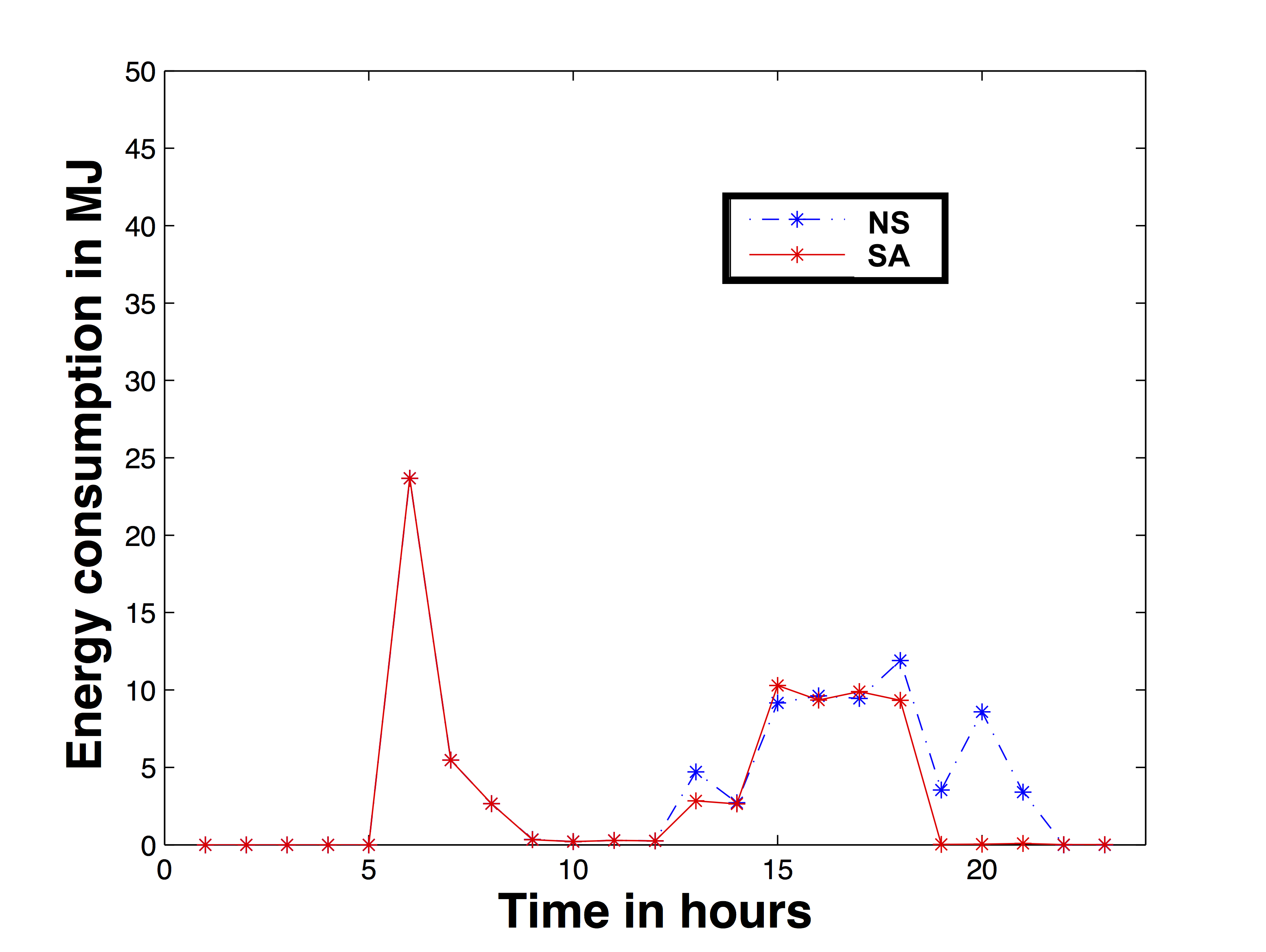}
     }
%     \hfill
     \subfloat[Scenario 3\label{subfig-32:dummy}]{%
       \includegraphics[width=0.33\textwidth]{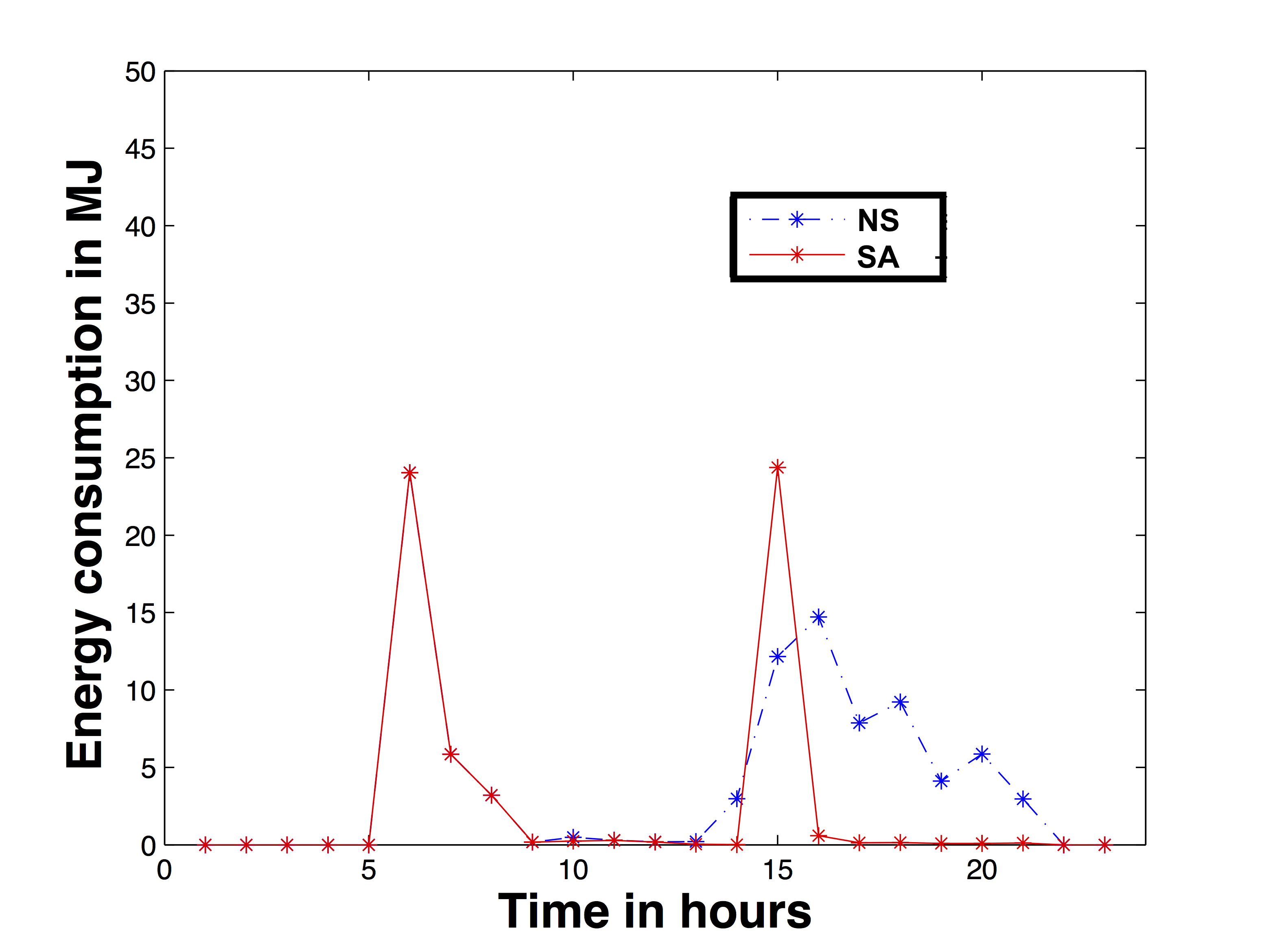}
     }
     \caption{Comparison of energy use by SA and NS for one typical \textbf{summer} day.}
     \label{fig:energy_summer}
\end{figure}

\begin{figure}[!ht]
	\subfloat[Winter\label{fig:winterplot}]{
		\includegraphics[width=0.5\textwidth]{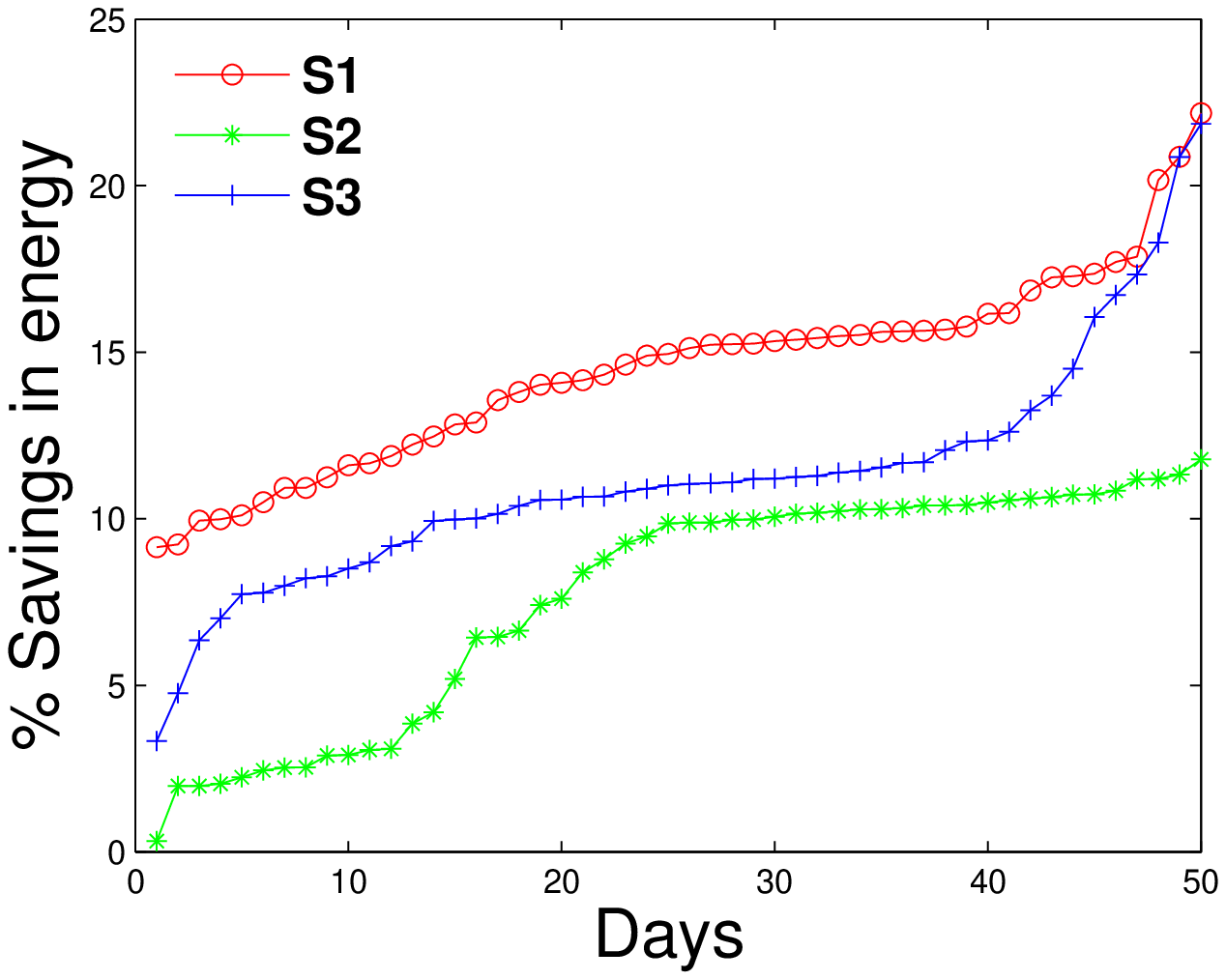}
 	}
 	\subfloat[Summer\label{fig:summerplot}]{
		\includegraphics[width=0.5\textwidth]{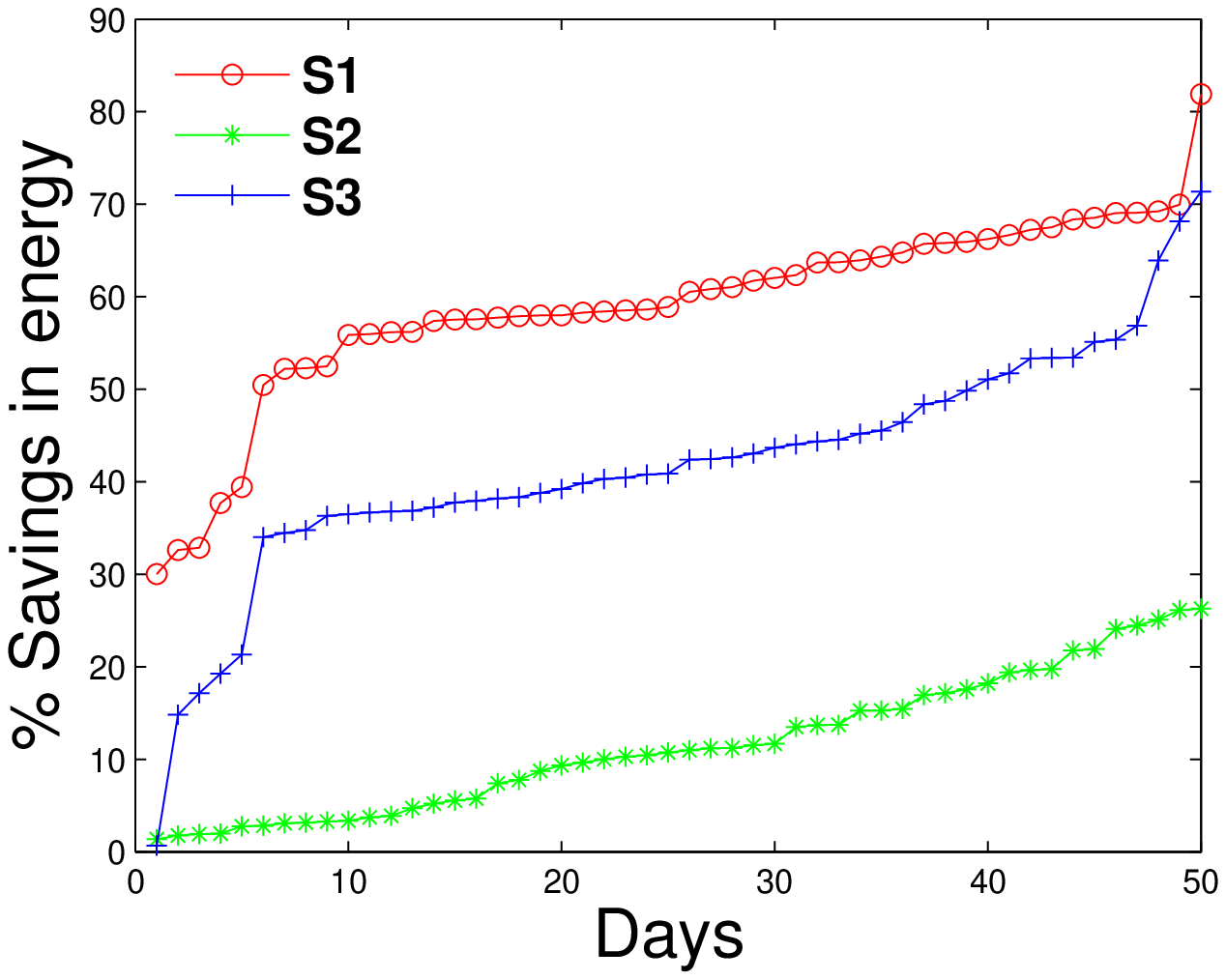}
    }
	\caption{Savings in energy when using SA instead of NS for 50 different days, sorted in ascending order for the three scenarios}
 	\label{fig:energy_savings}
\end{figure}

We  compare the energy consumed when using the SA system and  the NS system, for a typical summer day and a typical winter day. 
We observe that the energy consumption with SA, on a typical day, is
lower, as shown in Figures \ref{fig:energy_winter} and \ref{fig:energy_summer}.  
Note that the energy consumed by  SPOT-aware  HVAC at a specific time instant is not always lower than the energy consumed by NS. %but that is followed by a steep fall in the energy consumption. This is because an MPC  controller optimizes energy over a horizon as opposed to at one point of time.
Figure \ref{fig:winterplot} and Figure \ref{fig:summerplot} summarize the savings in energy in winter and summer respectively over 50 days. 
The average savings in energy, over the 50 days, for the three scenarios are reported in  Table \ref{table:energy}.

\begin{table}[ht]
\centering
 \begin{tabular}{|c|l|l|l|}
  \hline
\hline
Scenario &  Winter & Summer \\
\hline
1 & 14.44 & 59.02\\
 2 & 7.68& 11.62\\
 3 & 11.21& 41.88\\
\hline
\end{tabular}
\caption{Average percentage energy savings for S1, S2 and S3}\label{table:energy}
\end{table}
% Figure \ref{fig:S1_summer},  \ref{fig:S2_summer} and  \ref{fig:S3_summer} shows the comparison of the
% energy consumption with their respective benchmarks. 

We observe that in both seasons, there is a significant reduction in energy use 
with SA, compared to NS. 
In Scenario~1, this is because all the rooms have SPOT and hence there is more flexibility in the control process 
that uses the central HVAC only to
provide a base thermal level, 
with SPOT providing the additional thermal offset for any rooms
with occupancy.  

In Scenario~2, there is a room which does not have SPOT.
The central HVAC system is solely responsible for the thermal comfort in that room. 
Hence the occupancy pattern in
the common room is the primary determinant of the HVAC system operation: if this room is occupied,
the central HVAC needs to heat or cool all five rooms whether or not they are occupied.
As a result, we could not obtain the same savings in energy as with Scenario 1. 

In comparison to Scenario 2, we observe additional savings in energy in Scenario 3 because of 
the separation of the rooms with SPOT  and the rooms without SPOT into separate zones. 
Specifically, since all rooms in one of the zones have SPOT, there is more flexibility in the VAV control for this zone. 
Nevertheless, the AHU control is common to both zones, so we do not
obtain the same savings in energy as in Scenario 1.
This analysis suggests that to maximize energy savings,
SPOT should be deployed in all rooms in one entire zone at a time, rather than piecemeal.

When comparing energy savings across seasons,
we observe higher savings in energy  in summer than in winter.
This is because in summer SPOT cools with a fan, which needs only 30 W of power,
whereas in winter the SPOT heater consumes 700 W.
Hence, in winter if all the rooms were occupied, 
it is better to employ the central HVAC to 
provide the appropriate thermal level as opposed to operating the HVAC at a base level along with the SPOT systems in 
all the rooms being ON.
In other words, SPOT in heating mode is beneficial only during times of partial occupancy.
In contrast, in summer, even if all the rooms with SPOT were occupied, the rooms could be at a higher temperature than the desired level with the fan in all the rooms being ON, to maintain an
appropriate comfort level.

One of the significant reason for energy savings we obtain for SA is the reduced supply air temperature of the HVAC system in comparison with NS. This is illustrated in the next section.

%We observed that the thermal capacitance of the rooms influence the results. If the thermal
%capacitance was reduced then the savings in energy obtained also reduces.
\subsubsection{AHU Supply Air Temperature}

\begin{figure}[!ht]
     \subfloat[Scenario 1 \label{subfig-1:dummy}]{%
       \includegraphics[width=0.33\textwidth]{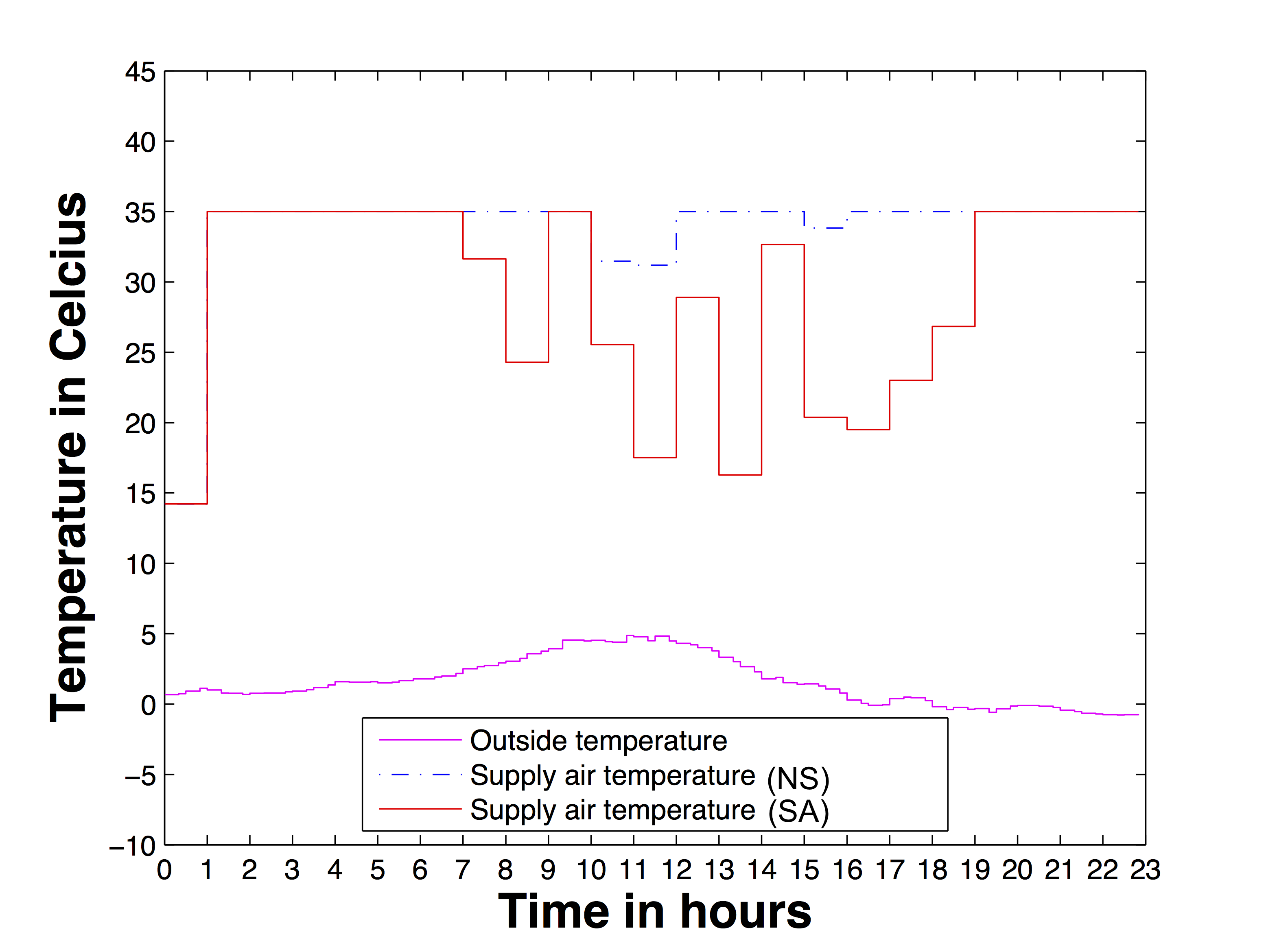}
     }
     \subfloat[Scenario 2\label{subfig-2:dummy}]{%
       \includegraphics[width=0.33\textwidth]{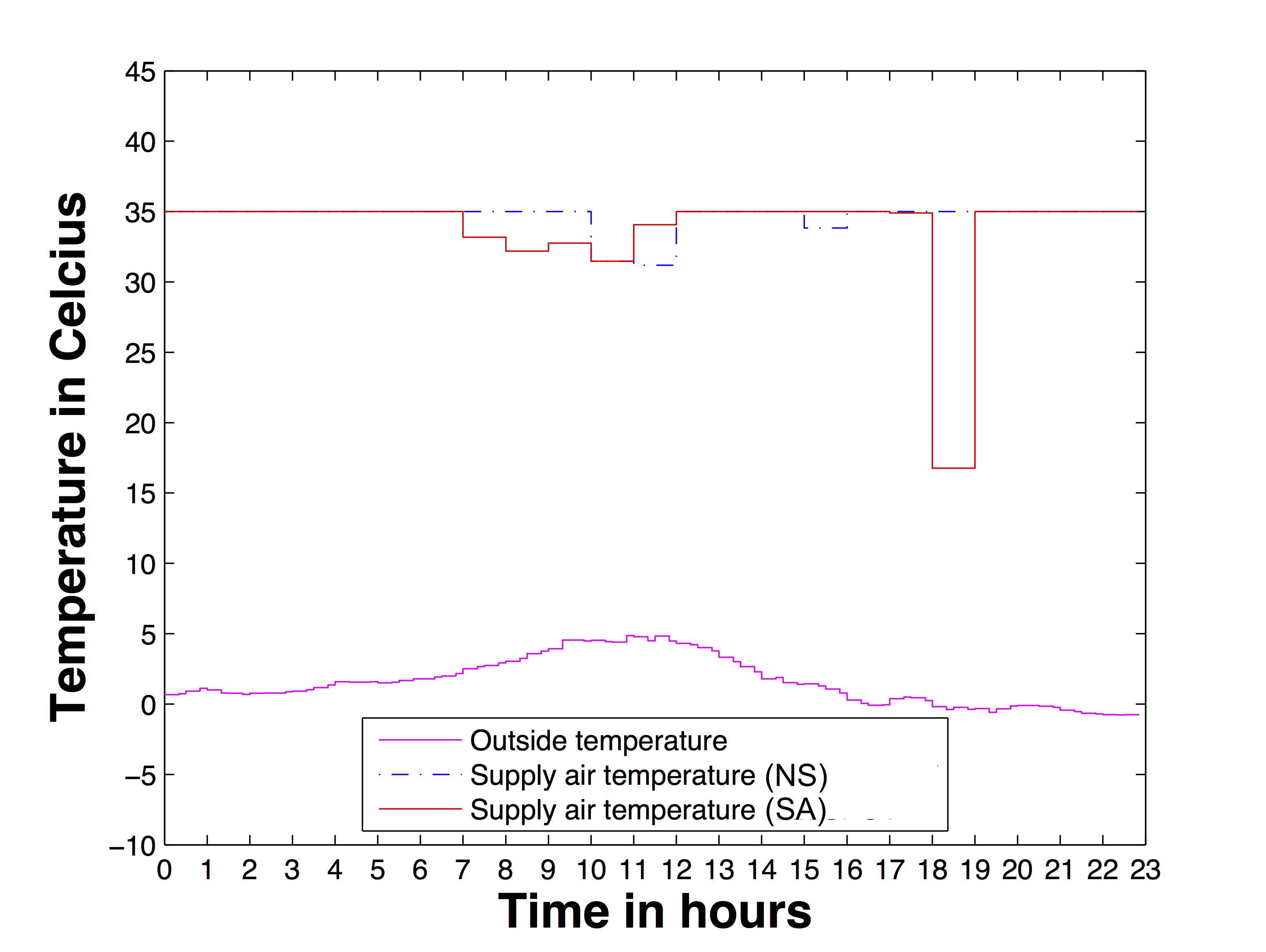}
     }
%     \hfill
     \subfloat[Scenario 3\label{subfig-2:dummy}]{%
       \includegraphics[width=0.33\textwidth]{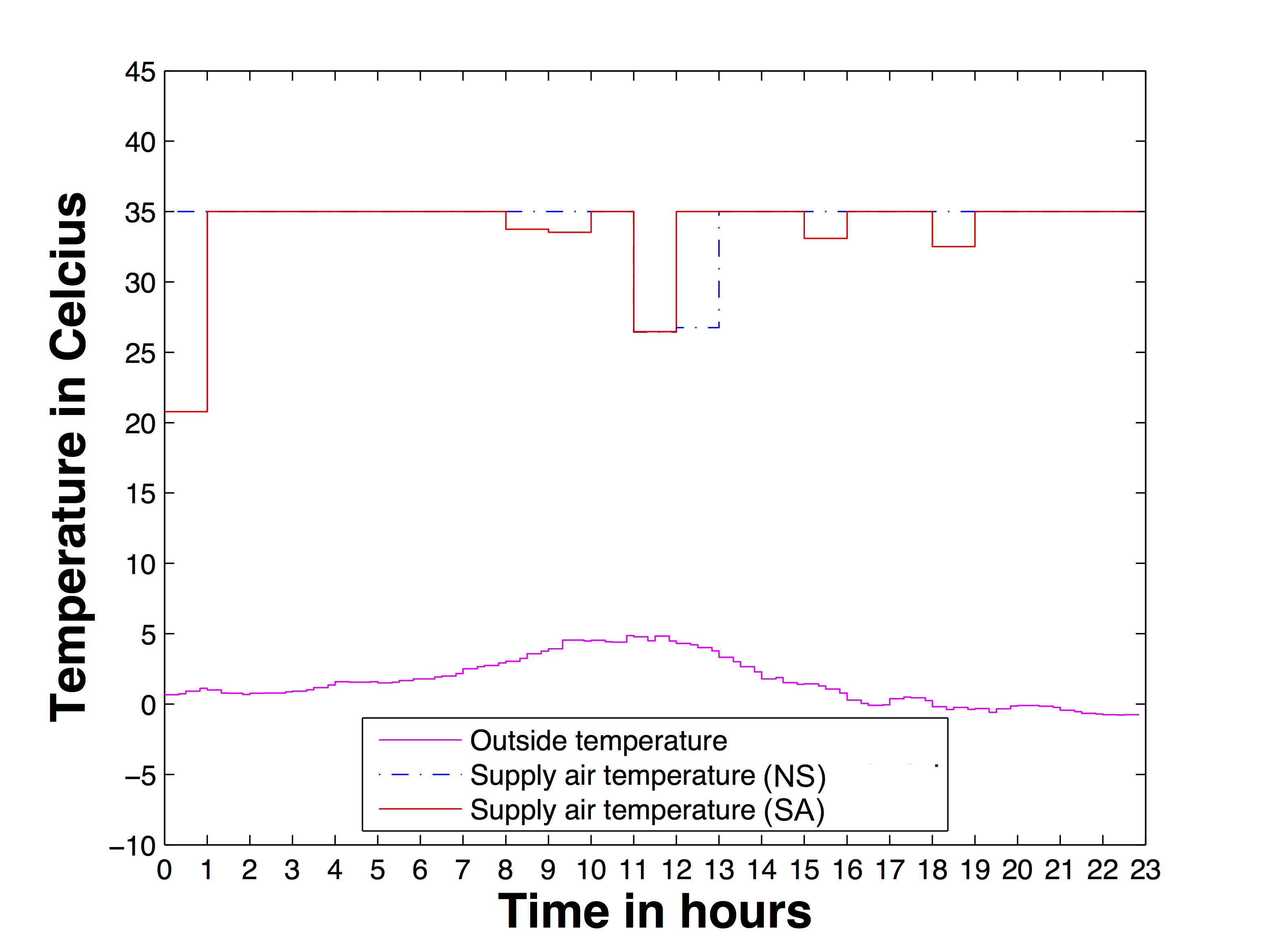}
     }
     \caption{Comparison of supply air temperature of SA and NS for one typical \textbf{winter} day.}
     \label{fig:DaySAT_winter}
\end{figure}

\begin{figure}[!ht]

     \subfloat[Scenario 1\label{subfig-1:dummy}]{%
       \includegraphics[width=0.33\textwidth]{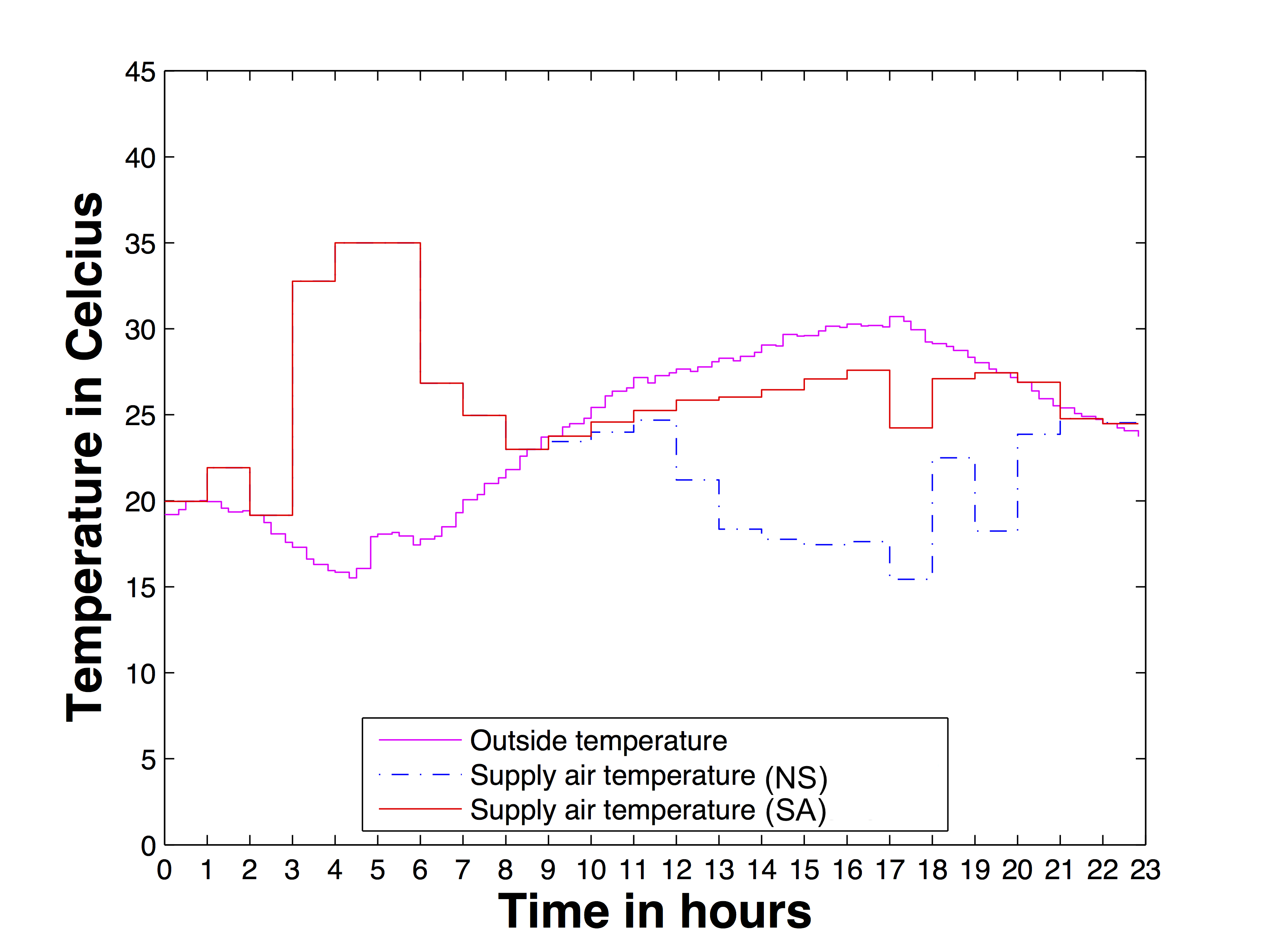}
     }
     \subfloat[Scenario 2\label{subfig-2:dummy}]{%
       \includegraphics[width=0.33\textwidth]{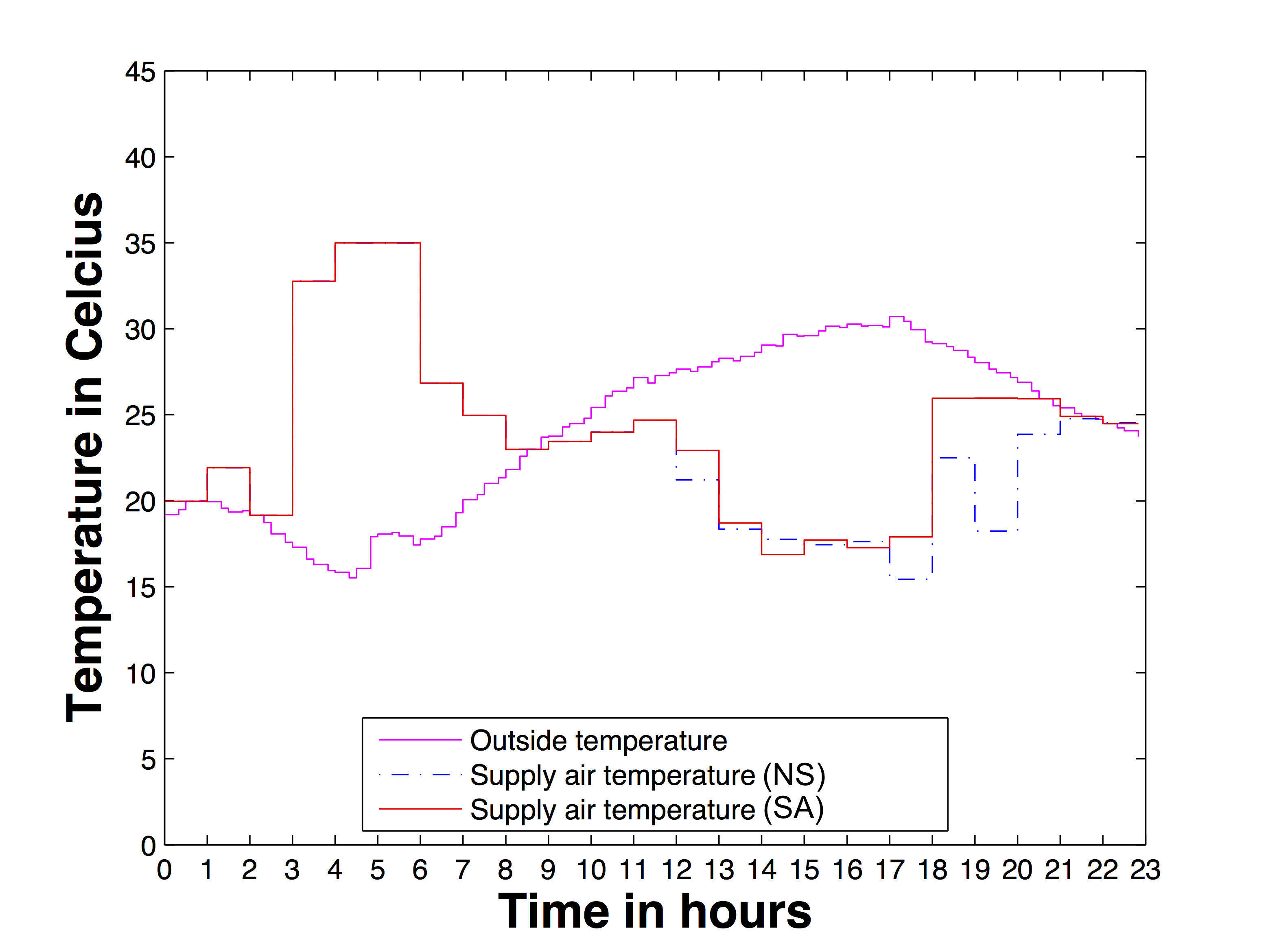}
     }
%     \hfill
     \subfloat[Scenario 3\label{subfig-2:dummy}]{%
       \includegraphics[width=0.33\textwidth]{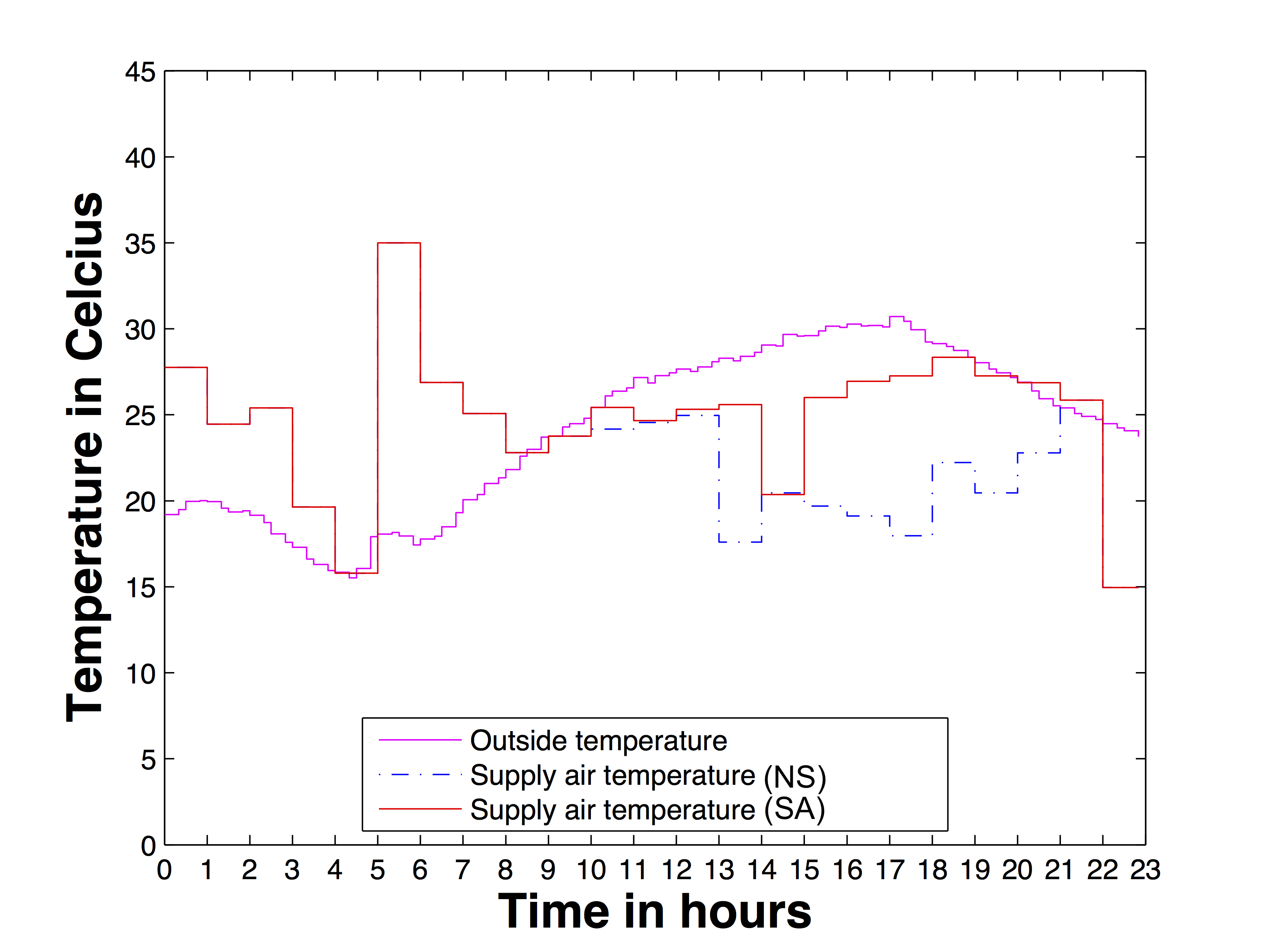}
     }
     \caption{Comparison of supply air temperature of SA and NS for one typical \textbf{summer} day.}
     \label{fig:DaySAT_summer}
\end{figure}

We expect that in Scenario 1, since all the rooms have SPOT, the supply air temperature could be lower (resp. higher) in  winter (resp. summer) with SA than NS. 
This is because SPOT would supply the additional offset in comfort.
Figure \ref{fig:DaySAT_winter} illustrates this for a single typical day in winter and
Figure \ref{fig:DaySAT_summer} for a typical day in summer.
   
We find that, as expected, the supply air temperature with SA is lower than with NS in winter, and higher in summer.
%For the 50 test days, the mean supply air temperature was \green{XXX degrees} lower in  winter
%and \green{YYY degrees} higher in summer in Scenario 1. 
%The corresponding figures for Scenarios 2 and 3 are 
%\green{fill this in, or better, make this into a table.}
We find that the greatest differences in supply
air temperature are with a full deployment of SPOT (S1), but there is a significant impact
even with partial deployments (S2 and S3).
\subsection{Heterogeneous Comfort Requirements}
To study the case of heterogeneous comfort requirements, for simplicity, we focus on Scenario 1,
that  is, a building with a single zone.
We assume that there are diverse comfort requirements across the rooms in this zone. 
These values, given in Table \ref{tab:pmvvalues}, are chosen close to the standard (ASHRAE) comfort requirements. Each user has comfort requirements that vary within a range of $1^{\circ}$C.
%\textcolor{green}{ Need to be more specific about how these values were chosen}

\begin{figure}[!ht]
	
    \subfloat[Energy\label{fig:winter_het_energy}]{
    	\includegraphics[width=0.5\textwidth]{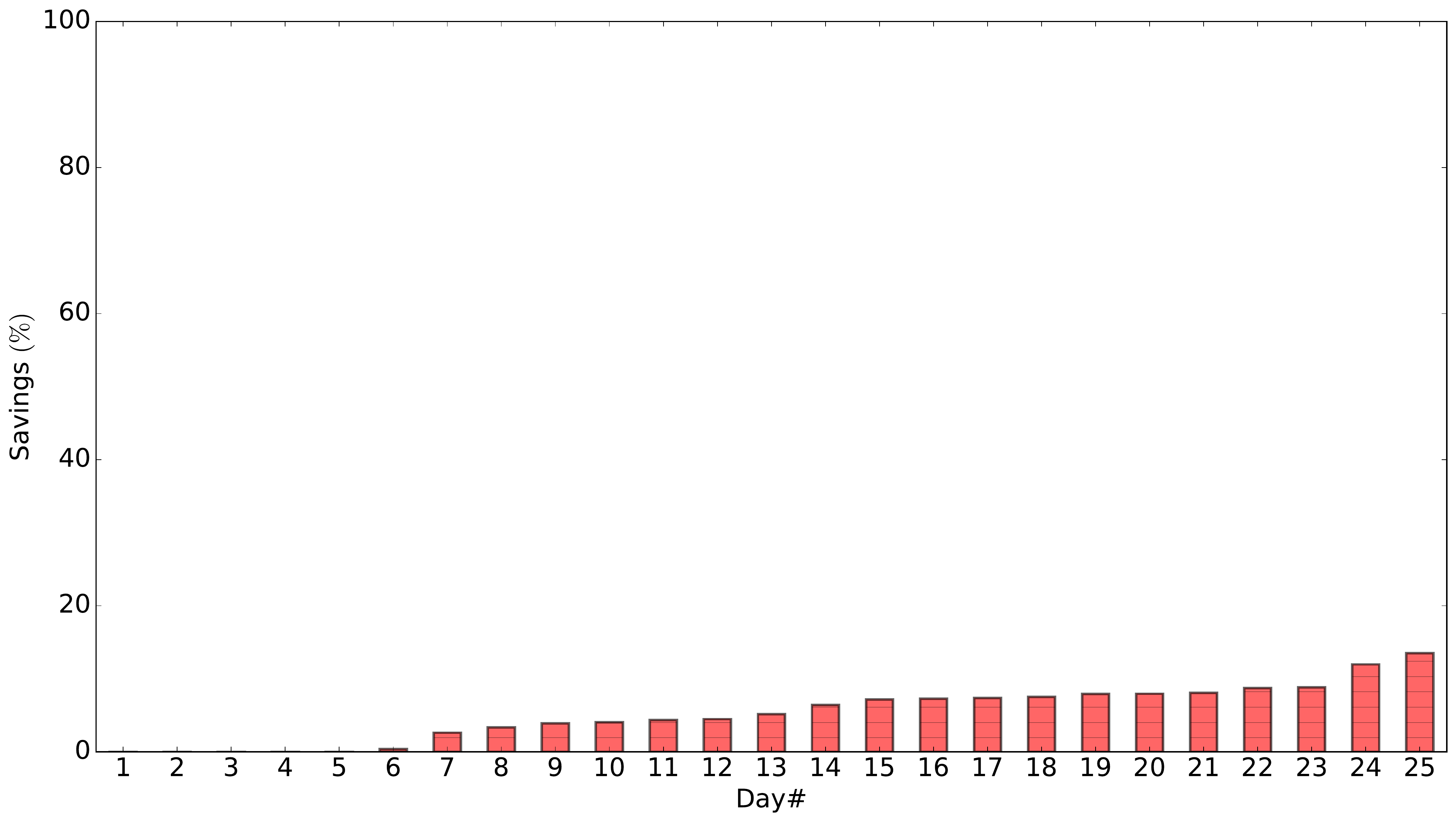}
	}    
    \subfloat[Discomfort \label{fig:winter_het_discomfort}]{
		\includegraphics[width=0.5\textwidth]{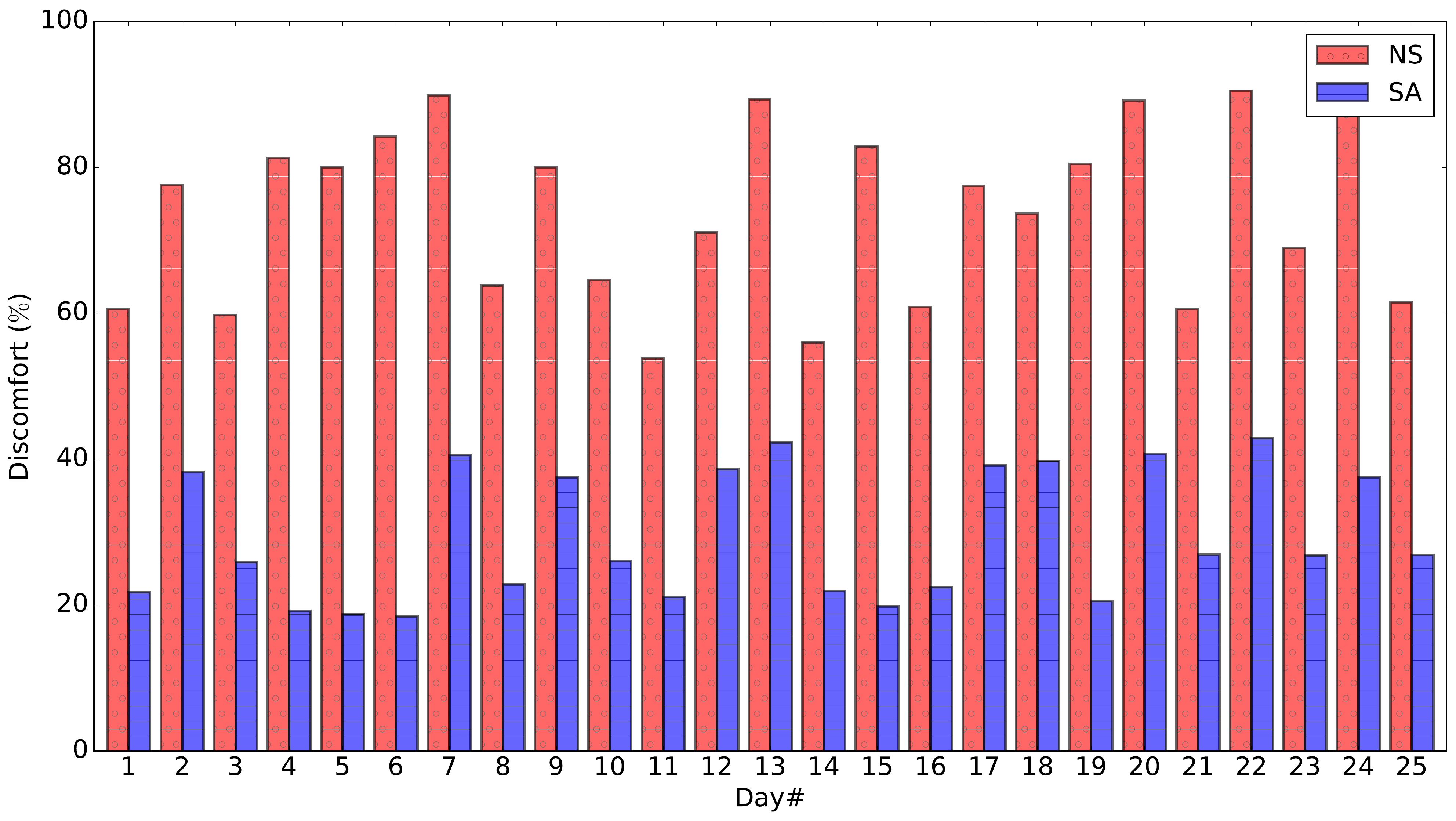}
    }
	\caption{Percent reduction in energy when using SA instead of NS (left) and average discomfort experienced by all occupants (right) for each control scheme  for 25 \textbf{winter} days.}	
	\label{fig:winter_het}
\end{figure}

\begin{figure}[!ht]

	\subfloat[Energy\label{fig:summer_het_energy}]{
       \includegraphics[width=0.5\textwidth]{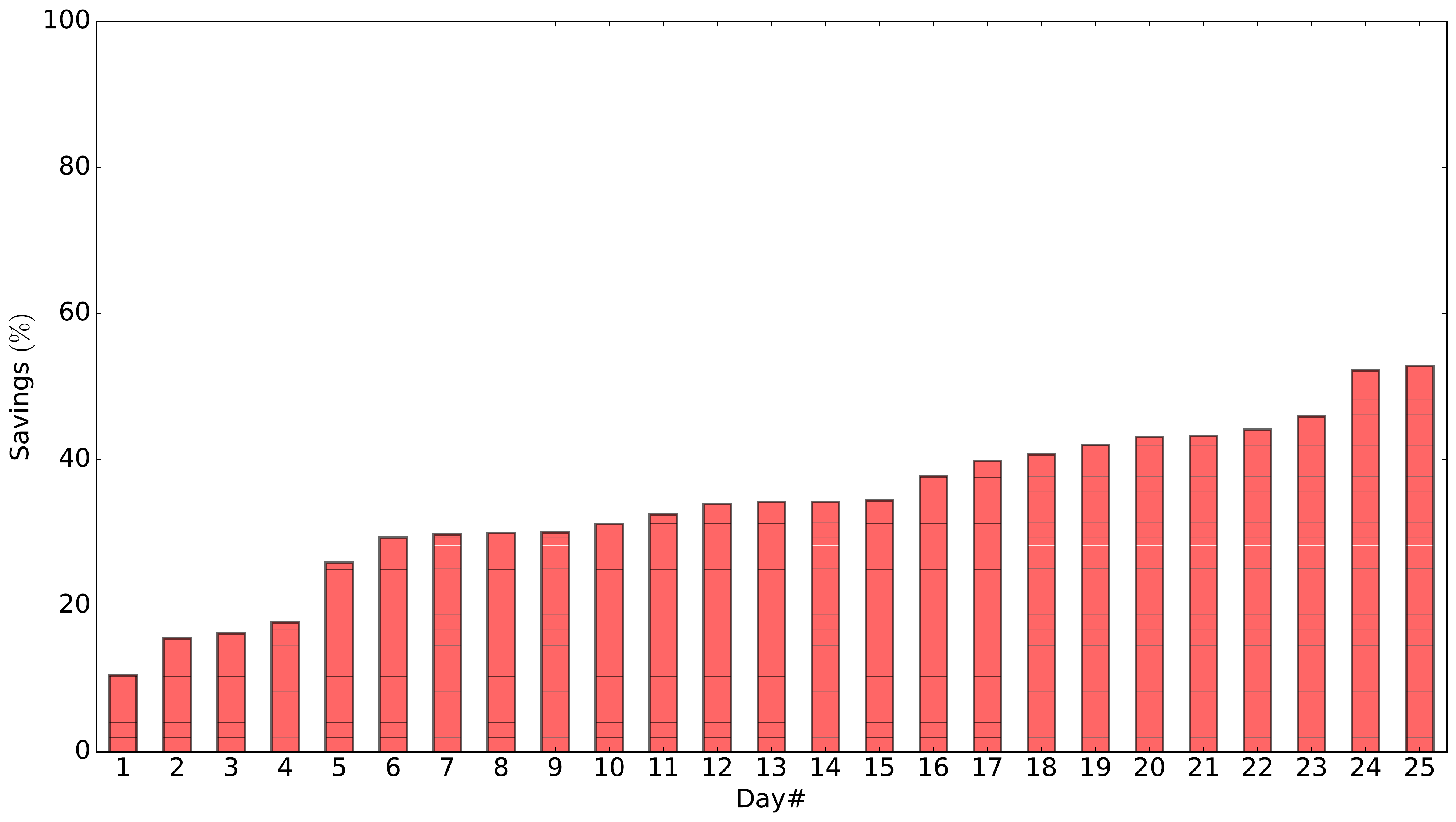}
     }
	\subfloat[Discomfort \label{fig:summer_het_discomfort}]{
       \includegraphics[width=0.5\textwidth]{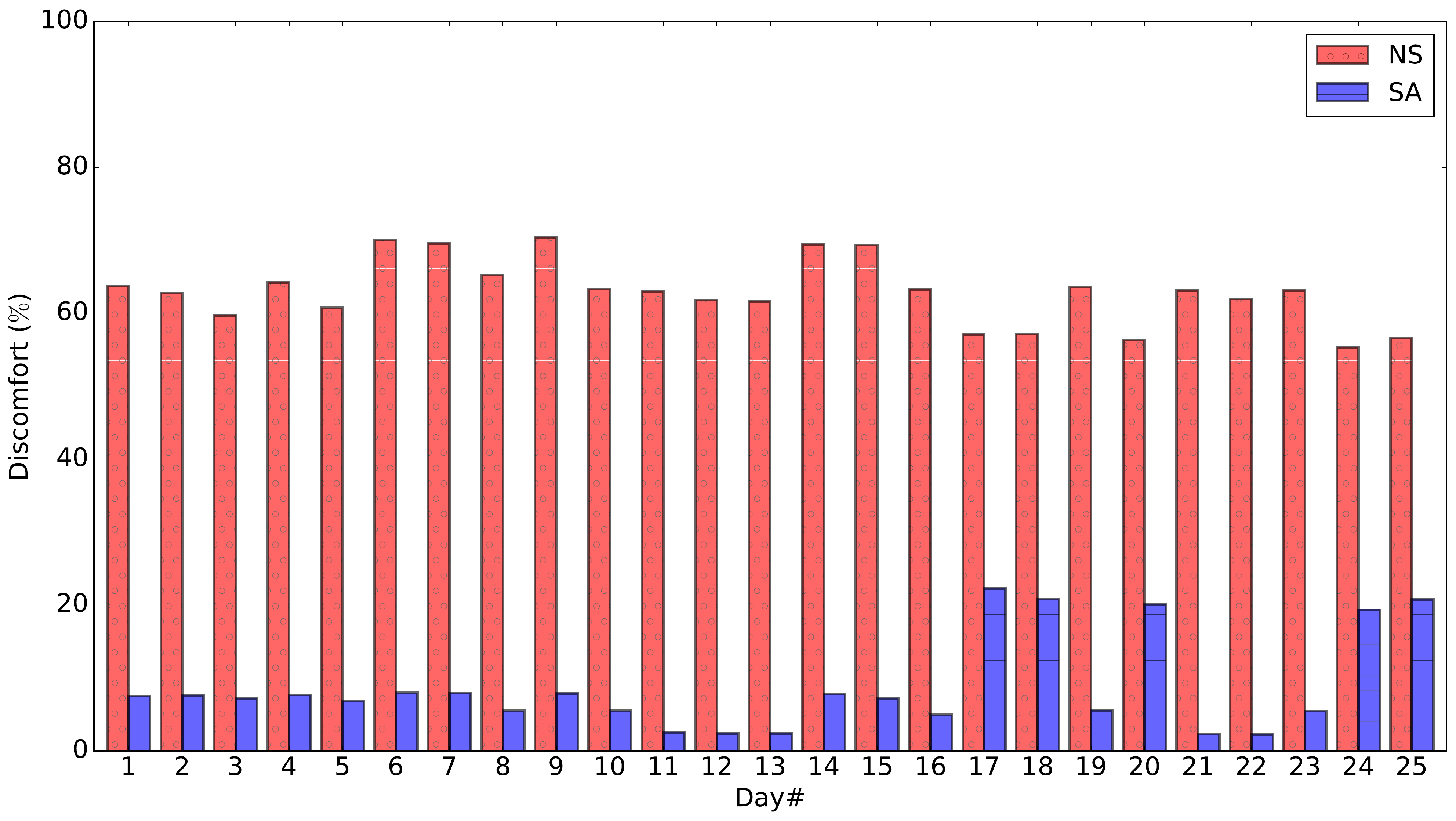}
     }
     \caption{Percent reduction in energy when using SA instead of NS (left) and average discomfort experienced by all occupants (right) for each control scheme  for 25 \textbf{summer} days.}	
     \label{fig:summer_het}

\end{figure}

With heterogeneous comfort requirements in a single zone, some users are likely to experience discomfort 
when using
NS. 
We employ the following metric to compute overall occupant discomfort. 
Since the smallest time-scale of control is 30 seconds, we consider every 30s interval in a day and check 
if a user experienced discomfort, i.e., the PMV level in the room lies outside the range specified by the user. 
We define
\begin{align*}
&\Delta_d(i)=  \mbox{Number of 30-second time intervals where user $i$ experienced discomfort in a day}\\
&\Delta_o(i) = \mbox{Number of 30-second time intervals the room was occupied by user $i$ in a day}\\
&D(i) =  \mbox{Discomfort experienced by user $i$ in a day} = \frac{\Delta_d(i)}{\Delta_o(i)} \\
&D = \mbox{Average discomfort experienced by all users in a day} = \frac{1}{N}\sum_{i=1} ^N D(i) \;  \; \mbox{($N$ is the number of rooms)}
\end{align*}
The average saving in energy consumed and  the average comfort improvement when using SA instead of NS over 25 days are reported  in 
Table \ref{tab:het_avgsaving}. 
Figure \ref{fig:winter_het_energy} and Figure \ref{fig:winter_het_discomfort} show the per-day results for winter while  Figure \ref{fig:summer_het_energy} and Figure \ref{fig:summer_het_discomfort} show the per-day results for summer.

 %\begin{tabular}{|l|c|c|}
  %\hline
%\hline
%Season &  \% savings in energy& \% improvement in comfort \\
%\hline
%Winter & 28.89& 50.41\\
%Summer & 68.91& 90.5\\
%\hline
%\end{tabular}
%\end{table}
\begin{table}[ht]\caption{Average (over 25 days) energy savings and improved comfort for the case with heterogeneous comfort requirements}\label{tab:het_avgsaving}
\centering
 \begin{tabular}{|l|c|c|}
  \hline
\hline
Season &  \% savings in energy& \% improvement in comfort \\
\hline
Winter & 32& 29\\
Summer & 82& 51\\
\hline
\end{tabular}
\end{table}

%Figure \ref{fig:comfort_winter} and Figure \ref{fig:comfort_summer} shows the per room comfort plots in winter and summer respectively for  S1 and B1 for one day. The
%corresponding occupancy pattern for both the plots is given in Figure \ref{fig:ocp1}
%in Appendix. Based on the occupancy the comfort requirements (upper and lower limits) are also in both Figure \ref{fig:comfort_winter} \& \ref{fig:comfort_summer}
%We observe that, in both seasons, for the benchmark system, the comfort 
 %constraints are often violated in all rooms except for Room 3. 
 %This is because all the rooms share a common control, which  
 %happens to match the requirements of the occupant(s) of Room 3. 
 %Occupants in the remaining rooms experience some discomfort.
 %Unlike B1, system S1, which has SPOT in all the rooms, is capable of providing diverse comfort requirements.
% \textcolor{green}{figure labels/legends need to be fixed}

\begin{figure}[!ht]
\subfloat[Summer\label{fig:summer_wwise}]{
       \includegraphics[width=0.5\textwidth]{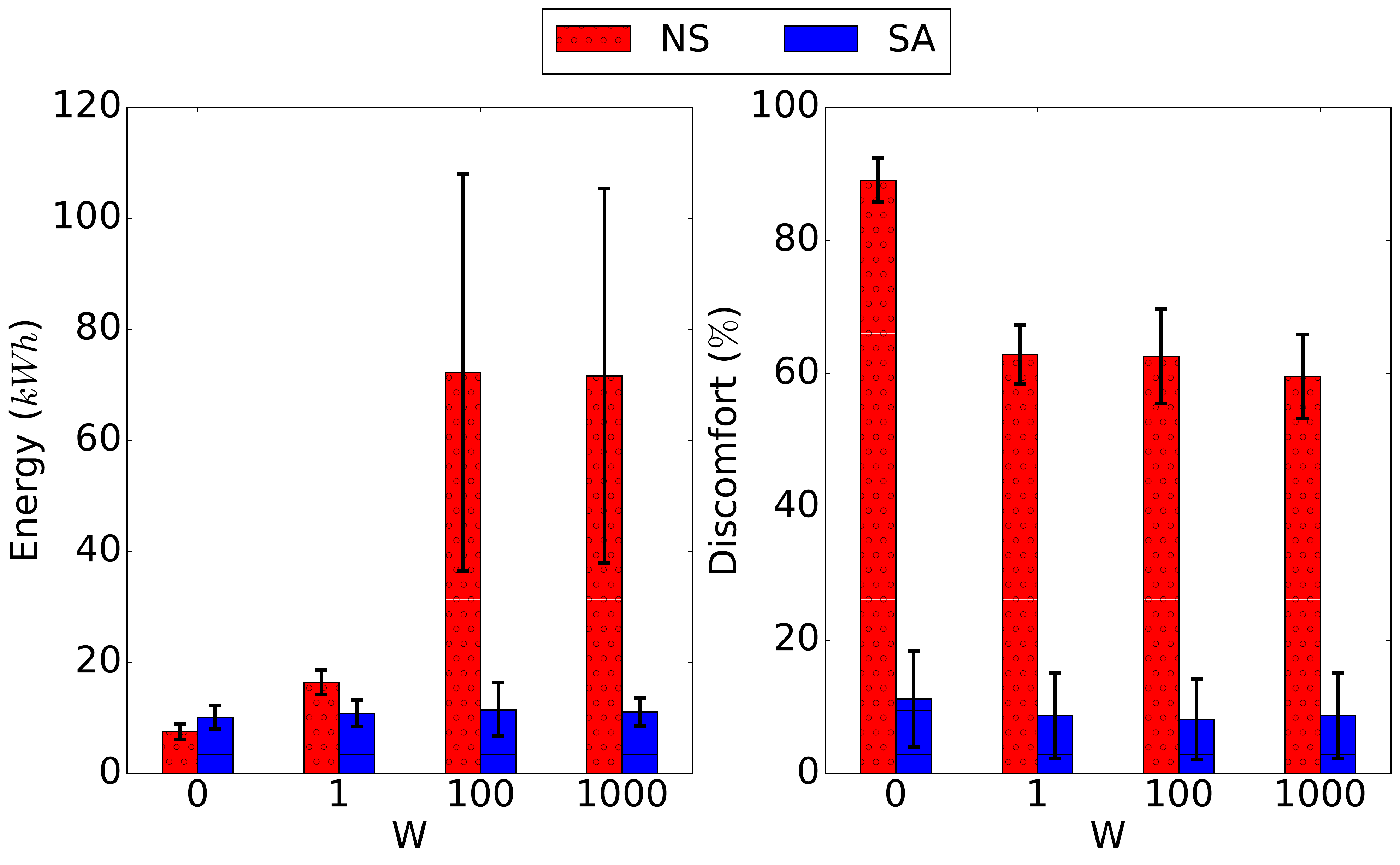}
     }
   \subfloat[Winter\label{fig:winter_wwise}]{
       \includegraphics[width=0.5\textwidth]{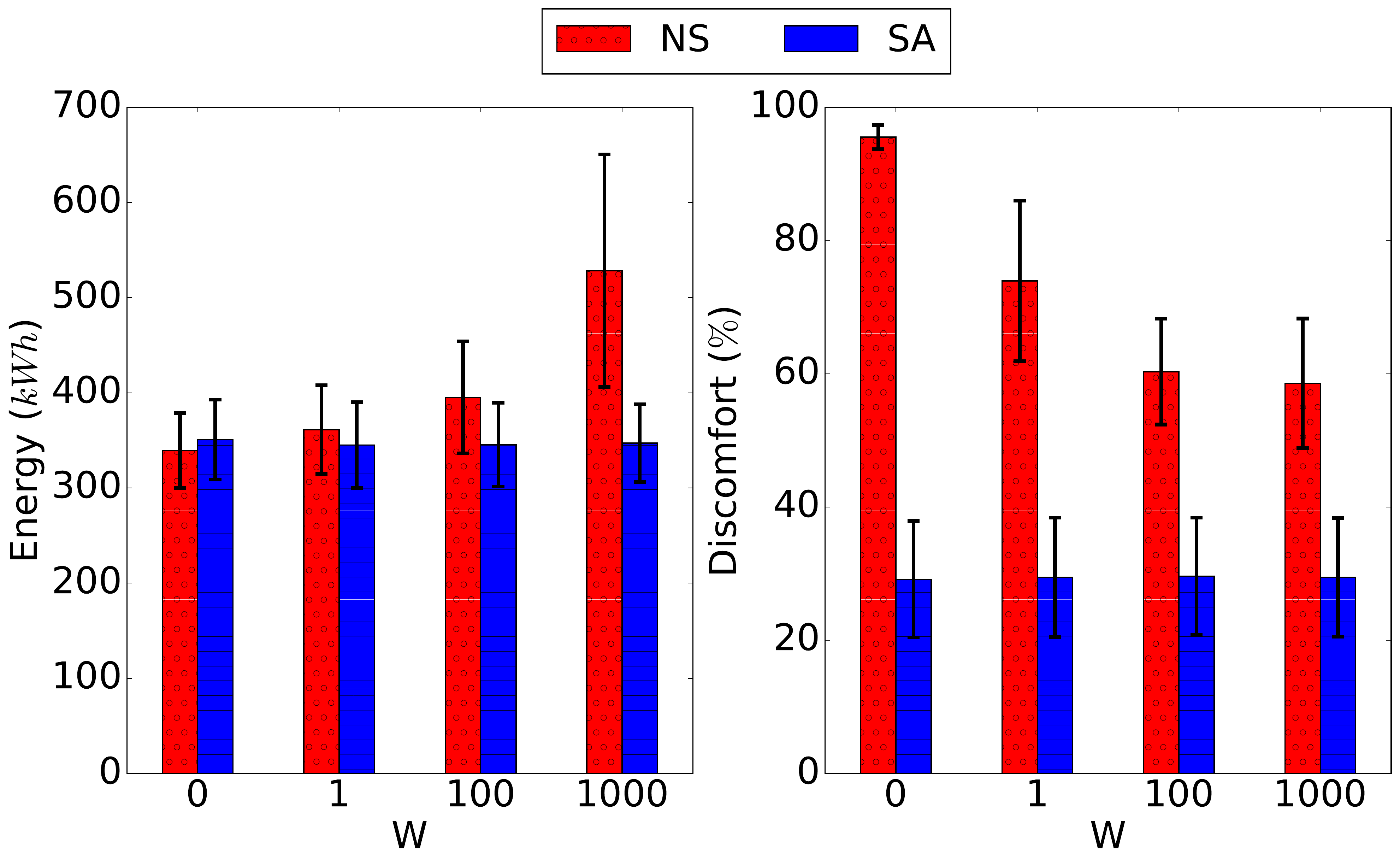}
     }
    
     	\caption{Comparing energy use and discomfort for NS and SA  for different values of the weighting parameter $W$.}	
     \label{fig:het_W_wise}
\end{figure}

A quick glance at Figures \ref{fig:winter_het_energy} and \ref{fig:summer_het_energy} shows that in both seasons,
SA brings a considerable reduction in energy use on most days. 
%In winter, the mean reduction in energy was
%\textcolor{green}{XXX\%} and in summer, the mean reduction was \textcolor{green}{YYY\%}.
Moreover, Figures \ref{fig:winter_het_discomfort} and \ref{fig:summer_het_discomfort} shows that in both seasons,
SA had overall lower discomfort than NS.
%: a mean discomfort of \textcolor{green}{XXX} for SA versus \textcolor{green}{YYY} for NS in summer, and a mean discomfort of \textcolor{green}{XXX} for SA versus \textcolor{green}{YYY} for NS in winter.
This is because there are heterogeneous requirements in the five rooms, which could not be satisfied by the central HVAC
alone.  
%Note that there is some discomfort for the system with SPOT in Winter. This is because when the heater of the SPOT is employed it takes some time for the temperature of the surrounding region to increase. Hence this results in some intervals of discomfort according to our metric which is reported in Figure \ref{fig:winter_het_discomfort}. However, when the heater is ON, the fan is also ON for safety purpose (to prevent the surrounding plastic from melting)\cite{rabkes16}. This results in hot air being blown around the occupant which makes an occupant feel comfortable even if the surrounding temperature is slightly outside the desired PMV specified by the user. Hence the discomfort experienced by the user in real time is much lesser than what is reported in Figure \ref{fig:winter_het_discomfort} for the SPOT system.

We now comment on some interesting aspects of these figures:
\begin{itemize}
\item In winter (Figure \ref{fig:winter_het_energy}),  some days see insignificant savings in energy with SA. 
Those were the days with full occupancy,
so that SPOT had to be ON in all the rooms, reducing the energy gains. Nevertheless, on the same days, there is still 
a significant reduction in discomfort (see Figure \ref{fig:winter_het_discomfort}).
\item In winter there is certain amount of discomfort even when using SA. This is because, when a room
becomes occupied, SPOT needs some time to heat up the space and bring the temperature to the desired level. This would result in few intervals of discomfort for the user. 
\item  In summer, with SA, from the moment the fan is ON, the user perceives comfort and PMV reduces immediately. 
This process is quicker than employing the heater during Winter. Hence, with SA,  with respect to comfort,
we observe better performance in summer than in winter. 
\item With SA, we still observe a small amount of discomfort in summer.
This is due to the fact that sometimes, a heater needs to be employed to satisfy some occupant's comfort requirement
(when the HVAC set point is lower than the occupant's comfort level). During the time that it takes to heat the room,
the user would experience discomfort. 
\end{itemize}

Figure \ref{fig:het_W_wise} summarizes the prior four figures and also shows the impact of the weighting value W in Equation \ref{eqn:addnterm}.
We see that the reduction in energy use and discomfort is greater in summer
than in winter.
Increasing the value of the weighting factor $W$ causes  NS
 to expend more energy to reduce discomfort. 
However, even with a large expenditure in energy,
NS is unable to match the performance of SA 
in either summer or winter, though in winter, the performance gap is smaller.
Specifically, with a small value of $W$, NS expends nearly the same energy
as SA. However, this increases discomfort well beyond what is achieved
by SA. Gains in comfort can only be achieved by expending more energy,
even so, the comfort achieved by NS is always less than that achieved by SA.

The above evaluation shows the benefits in terms of both energy reduction and improvement in comfort
of using a SPOT-aware HVAC system, as compared to a 
 HVAC system without SPOT.

\subsection{The Price of Unawareness}
Ideally, a central 
HVAC control system should 
be modified to take into account the deployment
of SPOT instances in the building as proposed in this paper.
However, we realize that, at least at the outset, 
this may not always be feasible. 
Hence, we now evaluate another control variant, dubbed \textbf{SU},
where SPOT is introduced in the same
rooms as in  SA,
but the central HVAC is unaware of the presence of SPOT, and hence chooses
the same AHU and VAV set points as NS.

\begin{figure}[!ht]
\subfloat[Summer\label{fig:summer_SU}]{
       \includegraphics[width=0.5\textwidth]{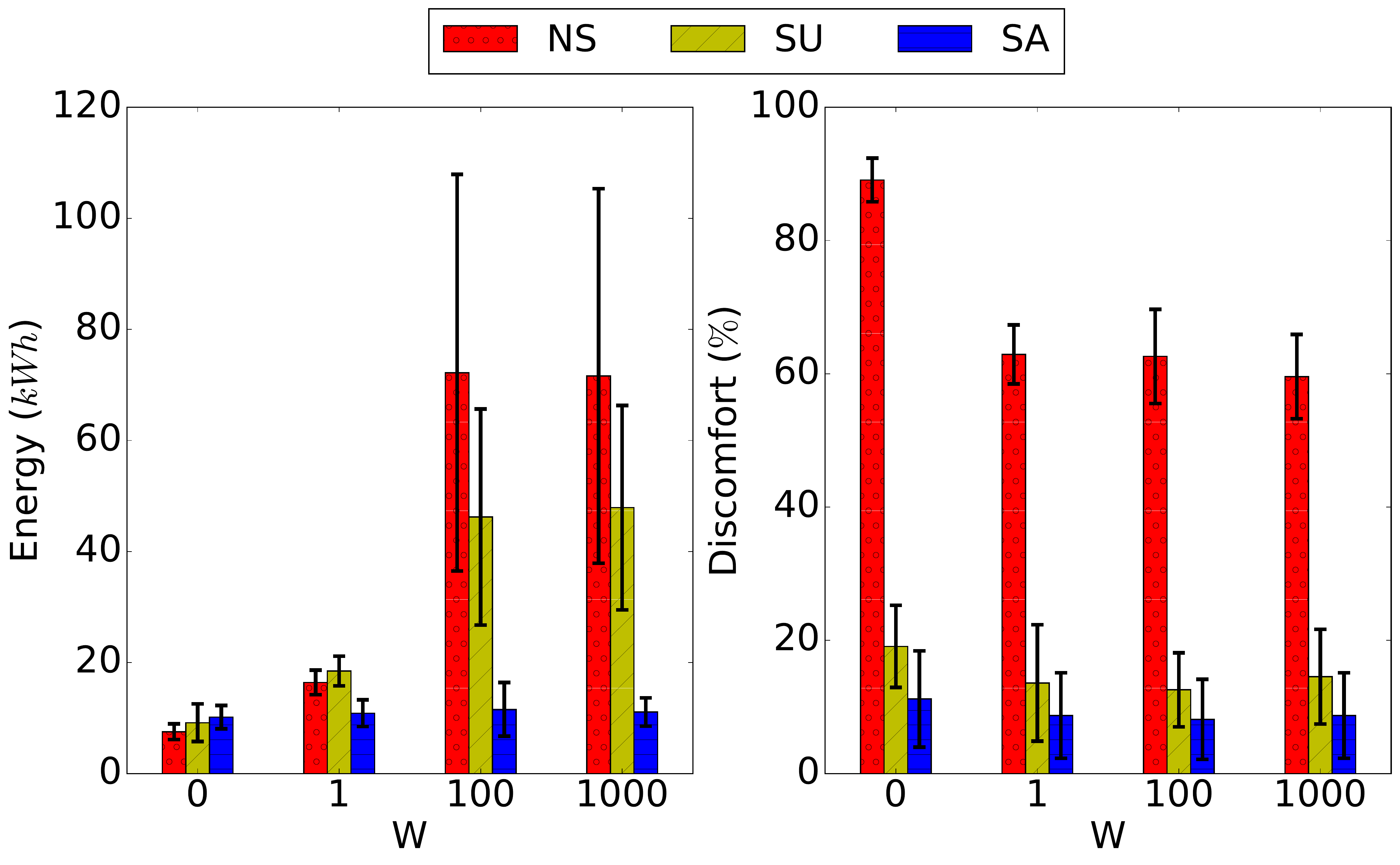}
     }
   \subfloat[Winter\label{fig:winter_SU}]{
       \includegraphics[width=0.5\textwidth]{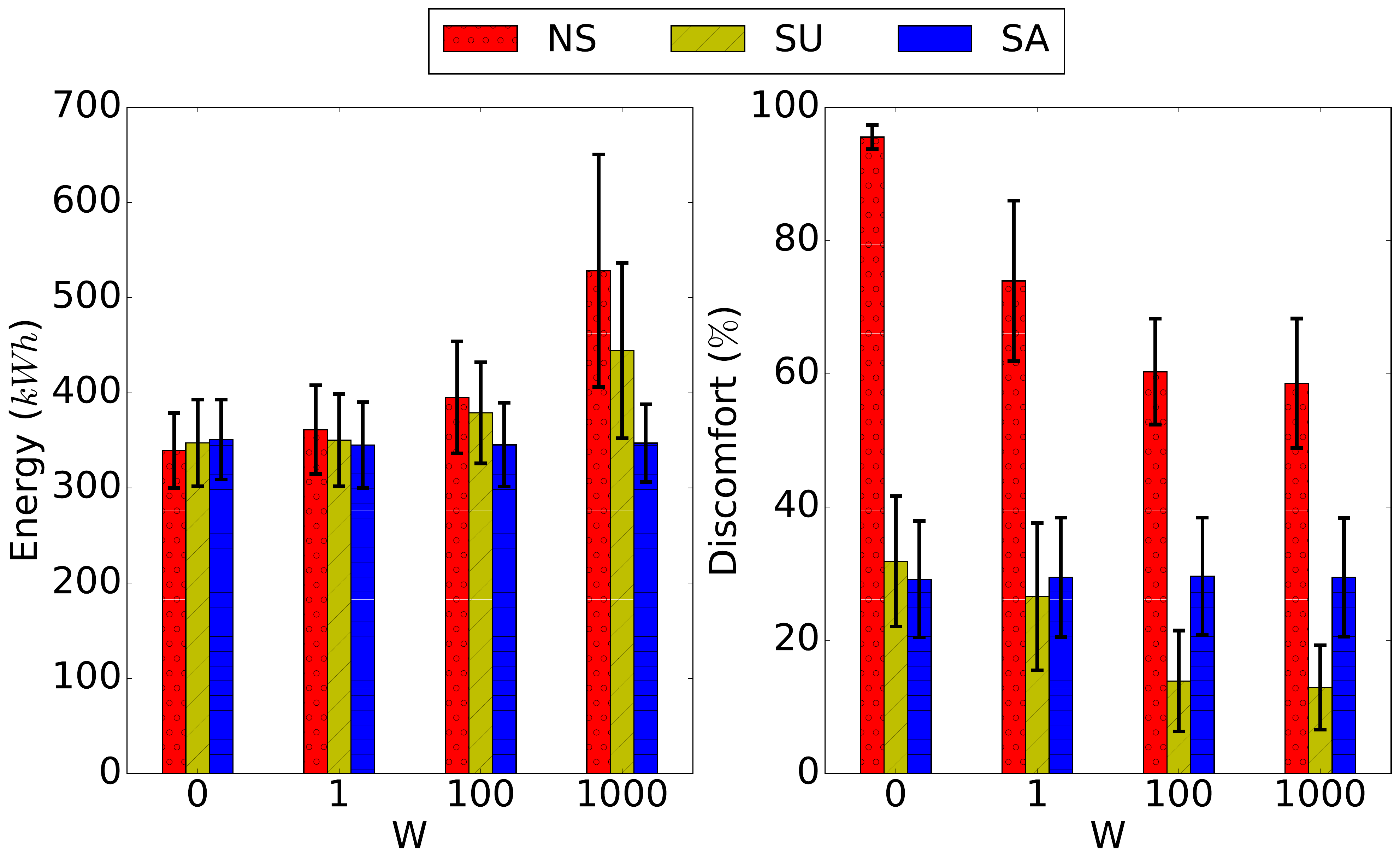}
     }
    
     	\caption{Comparing energy use and discomfort for SA, NS and SU systems for different values of the weighting parameter W.}	
     \label{fig:SU_W_wise}
\end{figure}

In the interests of space, we summarize the results for the 
SU scheme for Scenario 1 (full deployment of SPOT but
with partial occupancy) for heterogeneous comfort requirements
in Figure \ref{fig:SU_W_wise}. 

%\begin{figure}[!ht]
%	\includegraphics[width=\textwidth]{./figures/comb_pow_wday_div_SU}
    
%     	\caption{Energy use and discomfort for 25 \textbf{winter} days for  SA, NS and SU systems.}	
%     \label{fig:energy_winter}
%\end{figure}

%\begin{figure}[!ht]

     %\subfloat[Energy\label{fig:winter_het_energy_SU}]{
%       \includegraphics[width=0.5\textwidth]{./figures/pow_dw_wday_div_SU}
%     }
   % \subfloat[Discomfort\label{fig:winter_het_discomfort_SU}]{
%       \includegraphics[width=0.5\textwidth]{./figures/dcm_dw_wday_div_SU}
%     }
    
%     	\caption{Energy use and discomfort for 25 \textbf{winter} days for  SA, NS and SU systems.}	
%     \label{fig:energy_summer}
%\end{figure}

Note that SU does almost always much better than NS. The comparison of SA and SU is a little bit more challenging.
In summer, the discomfort achieved by SU
is statistically identical to that of SA, which is not
surprising, since they both have SPOT systems.
However, the energy expenditure with SU is significantly greater than
that of SA.

In winter, as $W$ increases,
the energy use of both  NS and SU  increases
and their discomfort reduces, as expected.
However, the discomfort of the NS system is significantly greater
than with the SU and SA systems.
The discomfort achieved by the SU system is lower than with the SA
system for large values of $W$ (though this
is not statistically significant). However, this comes at the cost
of increased energy use, compared to SA.
Essentially, this means that the SU controller makes a different
energy-comfort trade off than the SA controller.

 \section{Related Work}\label{sec:related}
 
A survey of personal comfort systems, quantifying their ability 
to provide comfort,
can be found in \cite{ZhaAreZha15}. In \cite{AttLer08}
a personal comfort system for cooling, essentially a fan working independent of the central cooling system, is proposed. 
Three versions of the SPOT  system are discussed in 
\cite{gaokes13,gaokes13a,rabkes16}. The first one uses many sensors and reactive control, the second proposes 
MPC-based optimal proactive control, the last one
uses a simple reactive control, and provides both heating (with a heating coil) and cooling (with a fan).
All three are designed to work independent of the central HVAC system.
 
The thermal model of a room is inherently bi-linear in nature \cite{kelmabor13}. A typical approach is to linearize the model about an operating point of the supply air temperature  and  then develop an optimal controller using methods such as Linear Quadratic Control (LQR) theory, or fuzzy logic. \cite{meh14,rahras03}. 
Since the linearization is done about an operating point, performance suffers when the operating 
point differs from the linearization set point. 
Consequently, we use a non-linear model, which avoids this problem.

For a non-linear model of the HVAC system with quadratic cost functions, feedback controllers are developed in \cite{SerRey99}. 
 In \cite{zhezah96}, the  control of 
 %\fix{[[one or multiple VAV?]]}
 a Variable Air Volume (VAV) unit of an HVAC system is formulated as  a non-linear optimization problem 
and solved using non-linear programming techniques. In both papers,
the optimization approach is myopic.  
%\fix{[[What do you mean by both the cases?]]}

Two comprehensive reviews of the role of MPC in HVAC systems can be found in \cite{mpcsurvey, kwazholi13}. 
We discuss only a few relevant papers here. 
\cite{maasan12} uses an MPC for minimizing the total and peak energy consumption of HVAC systems
in buildings. 
%Here a non-linear model is linearized about the set point temperature of the room. 
%For the internal load and external load, an affine model is developed and the parameters of the model are obtained from historical data. 
In \cite{AswMasTan12}, a hybrid system model is assumed for the HVAC system. The hybrid system is assumed to have certain number of modes,
with each mode having a fixed supply temperature, and
the model is assumed to be linear in each mode.  A new form of MPC called  Learning Based MPC (LBMPC)
is used.
%where the hybrid model is updated in order to account for the disturbances. 
In \cite{KelBor11}, a bi-linear system model is used and an MPC is employed. 
%The non-linear optimization
%is handled using the Sequential Quadratic Programming (SQP) approach.  
Here the operation of the HVAC 
system is modeled with a single time-scale. 
In \cite{GoyIngBar13} MPC-based control algorithms with special focus on occupancy information are proposed.
An analysis of the local 
optima for non-linear MPC for a single instance of the problem is done in 
\cite{kelmabor13}, where the influence of prediction horizon and discretization process on the local optima are investigated. 
Based on existing data from occupant feedback, a dynamic thermal comfort model was developed and used along with an MPC-based controller in \cite{CheWanSre16}. 
In \cite{CasAlvNorRod14}, non-linear MPC is used to determine the set points of an HVAC system,
which are then implemented using PID controllers. In \cite{FreOliGusNat08}, MPC-based control strategies are adopted to individually optimize thermal comfort and energy savings.  
Reference \cite{kalkesros16} is a preliminary paper where we investigate the idea of two time-scale HVAC control.

In this paper, we propose a novel approach to control 
a centralized HVAC system that is aware of comfort requirements
being met by the SPOT personal thermal comfort system. We
demonstrate that our approach results in significant savings in energy in addition to providing personalized comfort. To our knowledge, no prior work has considered the control
of a centralized HVAC system in the presence of such personal thermal comfort systems.

\section{Conclusion}\label{sec:conclusion}

Our work addresses two intrinsic problems of modern HVAC systems,
namely, coping with
diverse comfort requirements, and efficiently heating or cooling
partially occupied zones. 
We believe that using
personal thermal comfort systems, such as personal heaters or fans \cite{gaokes13,gaokes13a,rabkes16},
allows us to effectively
bridge the comfort gap between what is provided by a central 
HVAC system and the personal 
preferences of the occupants.
Thus,
we present the detailed design an MPC-based 
controller for a centralized HVAC system that is aware of
the deployment of
personal comfort systems and uses this knowledge to  exploit  
their responsiveness and flexibility for fine-grained thermal control.

Our control
algorithm explicitly models non-linearities in the physical system, 
resulting in a non-linear optimization problem.
It also explicitly models the physical constraints that limit the time-scales with 
which elements of
the central HVAC system can make changes to their set points, 
resulting in a two time-scale MPC control.

We conduct a detailed numerical evaluation of our approach
using realistic occupancy models, both in winter and in summer,
and with partial and full deployment of personal comfort systems,
along the twin axes of energy consumption and comfort.
We find that
that our system obtains, on average, 45\% savings  in  energy in summer, 
and 15\% in 
winter, compared with a state-of-the-art MPC controller, 
for the case when we assume homogeneous comfort requirements.  
For heterogeneous comfort requirements, we observe about 30\% improvement 
in comfort in winter and about 50\% in summer in addition to significant savings in energy.  
This validates our claims about the effectiveness of our approach. 

\section*{References}
 \bibliographystyle{model1-num-names}
\bibliography{elsarticle-template-1-num.bib}
\section*{Appendix}
In this section we list the numerical values of various parameters of our model that was used in our analysis. The optimization problem is also tabulated.

\begin{center}
\begin{longtable}[h]{|l|}
\caption{ Optimization problem formulation for the two time-scale MPC with multiple VAVs} \label{table:optprob} \\

\hline 
\multicolumn{1}{|c|}{To be computed at time $\ell=6p+q$, $p\in\{0,1,...\}$, $q\in\{0,1,..5\}$} \\ \hline 
\endfirsthead

\multicolumn{1}{c}%
{{\tablename\ \thetable{} -- continued from previous page}} \\
\hline
%\\ \hline 
\endhead

\hline \multicolumn{1}{|r|}{{Continued on next page}} \\ \hline
\endfoot

\hline \hline
\endlastfoot
$\I:=\{1,2,\dots,m\}$ (Set of zones)\\
$\Rc_1:=\{1,2,\dots,n_{1i}\}$ (Set of rooms of Type $S$ in zone $i$)\\
$\Rc_2:=\{n_{1i}+1,\dots,n_{2i}\}$ (Set of rooms of Type $\bar{S}$ in zone $i$)\\
$\T:=\{\ell,\ell+1,\dots,\ell+23\}$\\
 %$k:=t_0+k\tau$, where $\tau=10$ minutes\\
\hline
 {\bf  Given :} \\
Measured parameters at time $\ell$:\\
~~~Outside temperature and room occupancy $T_o(\ell)$, $\Oc_{ij}(\ell)$, $\forall j\in \Rc_1\cup \Rc_2$, $\forall i\in \I$\\
~~~Room temperature in region 1 and 2, $x^1_{ij}(\ell)$,$x^2_{ij}(\ell)$ $\forall j\in \Rc_1\cup \Rc_2$, $\forall i\in \I$.\\
% Forecasts (Outside temperatures and room occupancies):\\
% ~~~$T_o(k)$, $\Oc_i(k)$, $\forall k \in \T$, $\forall i\in \I$.\\
System parameters:\\
~~~ $A_{0_i}$, $A_{1_i}$, $B_{i}$,  $D_{1_i}$, $D_{2_i}$, $\tilde{A}_{0_i}$, $\tilde{B}_i$
$\underbar{$u$}$, $\bar{u}$, $\underbar{$v$}_i$, $\bar{v}_i$, $\bar{V}_a$, $\forall i\in \I$. \\
~~~ $\theta_1$, $\theta_2$, $\theta_3$, $\theta_4$, $\theta_5$,
  $f_1$, $f_2$, $f_3$, $f_4$.\\
Comfort parameters:\\
~~~$\underbar{$\gamma$}$, $\bar{\gamma}$, 
$\underbar{$\kappa$}$, $\bar{\kappa}$, 
$\underbar{$\beta$}_{ij}$, $\bar{\beta}_{ij}$, $\forall j\in \Rc_1\cup \Rc_2$, $\forall i\in \I$.\\
Input values from previous step (to be  used if $q\neq0$):
~~~$U$ \\
\hline
{\bf Objective:}\\
%\begin{tabular}{c}
$\underset{\{u(k),v(k),r(k),w_{ij}(k),v_{a_{ij}}(k), T_m(k), T_c(k), x^1_{ij}(k),x^2_{ij}(k),\Pm_{ij}(k)\}}{\mbox{ Minimize}}$ 
$\sum\limits_{k=\T}\J(k)\tau$, \\
%\end{tabular} \\
% $ \underset{\{(u(k)),(v(k)),(r(k)), (w(k)), (T_m(k)), (x_i(k)) %\}}{\mbox{ Minimize }}
%  \underset{k\in \T, i \in \I}{\sum}\J(k)$\\
where $
\J(k):=\underset{i \in \I}{\sum}v_i(k)\theta_1(u(k)-T_c(k))+\underset{i \in \I}{\sum}v_i(k)\theta_2(T_m(k)-T_c(k))
+\theta_3 (\underset{i \in \I}{\sum}v_i(k))^2 $\\
$\hspace*{4cm} + \theta_4 \underset{\substack{i\in \I,\\ j\in\Rc_{1_i}}} \sum w_{ij}(k)+ \theta_5 \underset{\substack{i\in \I,\\j\in\Rc_{1_i}}}{\sum}v_{a_{ij}}(k).$\\
\hline 
{\bf Constraints:} \\
$x_{ij}(k+1)=A_{0_i}(i,i)x_{ij}(k)+A_{1_i}(i,i)x_{ij}(k)v_i(k) + 
B_i(i)u(k)v_i(k) + D_{1_i}(i,i)(k)T_o(k)$\\
\hspace*{2cm}$ + D_{2_i}(i)(k)\Oc_{i}(k)$, $\forall j\in \Rc_1\cup\Rc_2$, $\forall i\in \I$, $\forall k\in \T$.\\
$\Delta x_{ij}(k+1) = \tilde{A}_{0_i}(i,i)\Delta x_{ij}(k)+\tilde{B}_i(i)w_{ij}(k)$, $\forall j\in \Rc_1$, $\forall i\in \I$, $\forall k\in \T$.\\
$x^1_{ij}(k+1)=x_{ij}(k+1)+\Delta x_{ij}(k+1)$, $\forall j\in \Rc_1$, $\forall i\in \I$, $\forall k\in \T$.\\
 $x^2_{ij}(k+1)=x_{ij}(k+1) + D_3\Delta x_{ij}(k)$, $\forall j\in \Rc_1$, $\forall i\in \I$, $\forall k\in \T$.\\
$T_{m}(k)=[r(k)\frac{1}{mn}\underset{i \in \I, j\in \Rc_1\cup \Rc_2 }{\sum}x_{ij}(k)]+(1-r(k))T_o(k)$,$\forall k\in \T$.\\
$P_{ij}(k+1) = f_1x_{ij}(k+1)+ f_2v_{a_{ij}}^2(k) +f_3v_{a_{ij}}(k)+f_4$,
$\forall j\in \Rc_1$, $\forall i\in \I$, $\forall k\in \T$.\\
$  \underbar{$\beta$}_{ij}\leq P_{ij}(k) \leq \bar{\beta}_{ij},\Oc_{ij}=1,\forall i\in\I,j\in\Rc_1$, $\forall k\in \T$.\\
$ \underbar{$\gamma$}\leq x^1_{ij}(k) \leq \bar{\gamma}, \Oc_{ij}=1,\forall i\in\I,j\in\Rc_1$, $\forall k\in \T$\\
$ \underbar{$\kappa$}\leq x_{ij}(k) \leq \bar{\kappa},\Oc_{ij}=1,\forall i\in\I,j\in\Rc_2$, $\forall k\in \T$.\\
$ \underbar{$\gamma$}\leq x_{ij}(k) \leq \bar{\gamma},\forall i\in\I,j\in\Rc_1\cup\Rc_2$,$\forall k\in \T$.\\
$ 0\leq w_{ij}(k)\leq 1$ $\forall j\in \Rc_1$ $\forall i\in \I$, $\forall k\in \T$.\\
$0\leq v_{a_{ij}}(k) \leq \bar{V}_a$ $\forall j\in \Rc_1$ $\forall i\in \I$, $\forall k\in \T$.\\
$w_{ij}(k)\leq O_{ij}(k)$ $\forall j\in \Rc_1$ $\forall i\in \I$, $\forall k\in \T$.\\
$v_{a_{ij}}(k)\leq O_{ij}(k)\bar{V}_a$ $\forall j\in \Rc_1$ $\forall i\in \I$, $\forall k\in \T$.\\
%$v_{a_{ij}}(k)\geq w_{ij}(k)\bar{V}_a$ $\forall j\in \Rc_1$ $\forall i\in \I$, $\forall k\in \T$. \\
$r(k)\leq 0.8$, $\forall k\in \T$. \\
$T_c(k)\leq T_m(k)$, $\forall k\in \T$.\\
$u(k)\geq T_c(k)$, $\forall k\in \T$.\\
$\underbar{$u$}\leq u(k) \leq \bar{u}$, $\forall k\in \T$.\\
$\underbar{$v$}\leq \underset{i \in \I}{\sum}v_i(k) \leq \bar{v}$, $\forall k\in \T$.\\
If $q=0$, $u(6z+\ell)=u(6z+j+\ell)$, $z\in\{0,1,2,3\}$, $j \in\{1,..,5\}$. \\
If $q\neq 0$, $u(\ell)=U$,\\
%$qu(1)=qU$, \\
$u(\ell)=u(\ell+j), j \in\{0,1,..,5-q\}$,\\
$u(\ell+6z-q)=u(\ell+6z-q+j), z\in\{1,2,3\}, j \in\{1,..,5\}$.\\
$u(\ell+6z-q)=u(\ell+6z-q+j),z=4,j \in\{0,1,..,q-1\}$.\\
\hline

\end{longtable}
\end{center}
The matrices in the thermal model are explained in Section \ref{subsec:type2model}and \ref{subsec:thermalmodel}. The values of $\theta_1$ and $\theta_2$ can be obtained using $\rho$, $\sigma$, $\eta_h$ and $\eta_c$ (Section \ref{subsec:optprob}). Refer Table \ref{table:parametervalue} for the parameter values. The values of $f_i$'s are available in Section \ref{subsec:pmv}
\begin{center}
\begin{longtable}{|l|l|l|l|}
\caption{Parameter Values} \label{table:parametervalue} \\

\hline \multicolumn{1}{|c|}{\textbf{Notation}} & \multicolumn{1}{c|}{\textbf{Description}} & \multicolumn{1}{c|}{\textbf{Value}}
& \multicolumn{1}{c|}{\textbf{Units}}
\\ \hline 
\endfirsthead

\multicolumn{4}{c}%
{{ \tablename\ \thetable{} -- continued from previous page}} \\
\hline \multicolumn{1}{|c|}{\textbf{Notation}} & \multicolumn{1}{c|}{\textbf{Description}} & \multicolumn{1}{c|}{\textbf{Value}}
& \multicolumn{1}{c|}{\textbf{Units}} \\ \hline 
\endhead

\hline \multicolumn{4}{|r|}{{Continued on next page}} \\ \hline
\endfoot

\hline \hline
\endlastfoot
$C$  &Thermal capacity of the room& 2000 & $kJ/K$\\
$\rho$& Density of air & 1.2041 & $kg/m^3$\\
$\sigma$& Specific heat of air & 1 & $kJ/(kg.K)$\\
$d$ &Internal load due to people/ equipments&  0.2& $kW$\\
%$Q_h$ Heat energy supplied by SPOT heater & & $kW$\\
 $\alpha_0$& Heat transfer coefficient between room and outside  & 0.048 & $kJ/(K.s)$\\

 $\eta_h$ &Efficiency of heating unit & 0.9 & - \\
$\eta_c$ &Efficiency of cooling unit & 0.9 & - \\
%$\eta_c$  & 4 & - \\ 
$\theta_3$ &Proportionality constant for  fan power consumption & 0.094 & $kW.s^2/kg^2$ \\
 $\theta_4$ &Power consumed by SPOT heater ($Q_h$) & 0.7 & $kW$\\
 $\theta_5$ & Power supplied by SPOT fan & 0.03 & $kW$\\
%$n$ &Number of rooms in the zone& 5 & \\
%  $N_1$&Horizon of MPC I & 4 & steps\\
%  $\tau_1$&Time step of MPC I & 1 & hour \\
%  $N_2$&Horizon of MPC II  & 5 & steps  \\
%  $\tau_2$&Time step of MPC II & 10  & minutes \\
% $N_3$ &Horizon of MPC III & 9 & steps  \\
%  $\tau_3$&Time step of MPC III & 1 & minutes \\
% $\underbar{$\beta$}$ &Lower limit of temperature & &\\
%& in room if occupied (summer)& 22 & $^{\circ}C$\\
%  $\bar{\beta}$ &Upper limit of temperature  & & \\
%& in room if occupied (summer) & 25 &$^{\circ}C$\\
%$\underbar{$\beta$}$ &Lower limit of temperature & &\\
%& in room if occupied (winter)& 21 & $^{\circ}C$\\
 % $\bar{\beta}$ &Upper limit of temperature  & & \\
%& in room if occupied (winter) & 23 & $^{\circ}C$\\
$\underbar{$\beta$}$ & PMV\footnote{The PMV limits correspond to temperatures 21$^{\circ}C$ and 23$^{\circ}C$ (winter) and 23$^{\circ}C$ and 25$^{\circ}C$ (summer).} lower limit in SPOT region for winter (summer)\footnote{Homogeneous comfort requirement case.} & -0.29 (-0.7)  & $-$\\
  $\bar{\beta}$ & PMV upper limit in SPOT region for winter (summer) & 0.21 (0) & $-$\\
  $\underbar{$\gamma$}$ &Room temperature lower limit non-SPOT region& 18 & $^{\circ}C$\\
  $\bar{\gamma}$ &Room temperature upper limit non-SPOT region & 28 & $^{\circ}C$\\
  $\underbar{$\kappa$}$ &Temperature lower limit in Type $\bar{S}$ room for winter (summer) & 21 (23) & $^{\circ}C$\\
  $\bar{\kappa}$ &Temperature upper limit in Type $\bar{S}$ room for winter (summer)& 23 (25) & $^{\circ}C$\\
 $\tilde{C}$  &Thermal capacity of region 1 in Type $S$ room & 200 & $kJ/K$\\ 
 $\alpha_r$& Heat transfer coefficient between & &\\
& two regions in Type $S$ room  &0.1425  & $kJ/(K.s)$\\
%$\underbar{$\gamma$}$ &Room temperature lower limit (unoccupied)& 18 & $^{\circ}C$\\
%  $\bar{\gamma}$ &Room temperature upper limit (unoccupied) & 28 & $^{\circ}C$\\
  $\underbar{$u$}$&Lower limit of $u$& 12 & $^{\circ}C$\\
  $\bar{u}$&Upper limit of $u$& 30 & $^{\circ}C$\\
  $\underbar{$v$}$&Lower limit of $v$ &0.236& $m^3/s$\\
  $\bar{v}$&Upper limit of $v$ &4.5 & $m^3/s$\\
$W$ & Weighting factor for discomfort & 1000 & -\\
\hline
\end{longtable}
\end{center}
 %\begin{figure}[H]
 %\includegraphics[width=\textwidth,center]{ocp1}
 %\caption{Occupancy pattern for a day in 5 rooms in a zone}\label{fig:ocp1}
%\end{figure}

\begin{table}[h]\caption{Desired temperature and PMV ranges for winter and summer for the case with homogeneous comfort requirements. The PMV values are obtained from PMV equations in Section \ref{subsec:pmv} by substituting the corresponding temperature values and fan speed 0.}\label{tab:pmvvalues:hom}
\center
 \begin{tabular}{|l|c|c|c|c|}
  \hline
  Parameter & \multicolumn{2}{|c|}{Winter} &  \multicolumn{2}{|c|}{Summer}\\
 \cline{2-5}
& Lower & Upper & Lower & Upper\\
 \hline
 Temperature&21 & 23& 23 & 25 \\
 \hline
PMV&-0.29 & 0.21& -0.7 & 0 \\
\hline
\end{tabular}
\end{table}
\begin{table}[h]\caption{Desired PMV ranges for winter and summer for the case with heterogeneous comfort requirements}\label{tab:pmvvalues}
\center
 \begin{tabular}{|c|c|c|c|c|}
  \hline
    Room& \multicolumn{2}{|c|}{Winter} &  \multicolumn{2}{|c|}{Summer}\\
 \cline{2-5}
 & Lower & Upper & Lower & Upper\\
 \hline
1 &-0.4 & -0.16 & -0.92 & -0.56 \\
\hline
2 & -0.29 & -0.04& -0.74 & -0.37 \\
\hline
3 &-0.16 & 0.08& -0.56 & -0.19\\
\hline
4 &-0.04 &0.21 &-0.37 & 0\\
\hline
5 &0.08 & 0.33 & -0.19 & 0.18\\
\hline
\end{tabular}
\end{table}

%pmv_l=[-0.64;-0.41;-0.16;0.08;0.33];
 %pmv_u=[-0.41;-0.16;0.08;0.33;0.58];
 
 %pmv_l=[-1.29;-0.92;-0.56;-0.19;0.18];
 %pmv_u=[-0.92;-0.56;-0.19;0.18;0.55];

\end{document}